%% file: HIN-13-005_temp.tex
\pdfoutput=1

\documentclass[11pt,twoside,a4paper,cmspaper,final,collab]{cms-tdr}
\def\svnVersion{428500P}\def\cmsCernNoTag{CERN-EP-2016-217}\def\cmsCernDate{\today}\def\cmsMessage{Published in Physical Review C as \href{http://dx.doi.org/10.1103/PhysRevC.96.015202}{\doi{10.1103/PhysRevC.96.015202.}}}

\begin{document}\cmsNoteHeader{HIN-13-005}

\hyphenation{had-ron-i-za-tion}
\hyphenation{cal-or-i-me-ter}
\hyphenation{de-vices}
\RCS$Revision: 416276 $
\RCS$HeadURL: svn+ssh://svn.cern.ch/reps/tdr2/papers/HIN-13-005/trunk/HIN-13-005.tex $
\RCS$Id: HIN-13-005.tex 416276 2017-07-14 14:12:04Z rkunnawa $
\providecommand {\rootsNN}  {\ensuremath{\sqrt{\smash[b]{s_{_{\mathrm{NN}}}}}}\xspace}
\newcommand {\npart}    {\ensuremath{N_\text{part}}\xspace}
\newcommand {\ncoll}    {\ensuremath{N_\text{coll}}\xspace}
\newcommand {\TAA}        {\ensuremath{T_\mathrm{AA}}\xspace}
\newcommand {\ak}    {anti-\kt\xspace}
\newcommand{\RR}{\ensuremath{R}\xspace}
\newcommand {\RAA}   {\ensuremath{\RR_{\mathrm{AA}}}\xspace}
\newcommand {\pp}    {\ensuremath{\mathrm{pp}}\xspace}
\newcommand {\PbPb}  {\ensuremath{\mathrm{PbPb}}\xspace}
\ifthenelse{\boolean{cms@external}}{\providecommand{\NA}{\ensuremath{\cdots}\xspace}}{\providecommand{\NA}{\ensuremath{\text{---}}\xspace}}
\providecommand{\HYDJET}{\textsc{hydjet}\xspace}
\providecommand{\PYTHJET}{\PYTHIA{}+\HYDJET}

\newlength\cmsFigWidth
\ifthenelse{\boolean{cms@external}}{\setlength\cmsFigWidth{0.98\columnwidth}}{\setlength\cmsFigWidth{0.7\textwidth}}
\ifthenelse{\boolean{cms@external}}{\providecommand{\cmsLeft}{top\xspace}}{\providecommand{\cmsLeft}{left\xspace}}
\ifthenelse{\boolean{cms@external}}{\providecommand{\cmsRight}{bottom\xspace}}{\providecommand{\cmsRight}{right\xspace}}
\ifthenelse{\boolean{cms@external}}{\providecommand{\cmsTable}[1]{#1}}{\providecommand{\cmsTable}[1]{\resizebox{\textwidth}{!}{#1}}}
\cmsNoteHeader{HIN-13-005}
\title{Measurement of inclusive jet cross sections in pp and PbPb collisions at \texorpdfstring{$\rootsNN=2.76$\TeV}{sqrt(s[NN])=0.276 TeV}}

\date{\today}

\abstract{Inclusive jet spectra from pp and PbPb collisions at a nucleon-nucleon center-of-mass energy of 2.76\TeV, collected with the CMS detector at the LHC, are presented. Jets are reconstructed with three different distance parameters (R = 0.2, 0.3 and 0.4) for transverse momentum (\pt) greater than 70\GeVc and pseudorapidity $\abs{\eta}< 2$. Next-to-leading-order quantum chromodynamic calculations with non-perturbative corrections are found to over-predict jet production cross sections in pp for small distance parameters. The jet nuclear modification factors for PbPb compared to pp collisions, show a steady decrease from peripheral to central events, along with a weak dependence on the jet \pt. They are found to be independent of the distance parameter in the measured kinematic range.}

\hypersetup{%
pdfauthor={CMS Collaboration},%
pdftitle={Measurement of inclusive jet cross-sections in pp and PbPb collisions at sqrt(s[NN])=2.76 TeV},%
pdfsubject={CMS},%
pdfkeywords={CMS, physics, heavy ions, jet quenching}}

\maketitle
\section{Introduction}

Heavy ion collisions at the CERN LHC can generate a hot and dense deconfined state of matter, also known as the quark-gluon-plasma (QGP). In these collisions, hard scattered partons are expected to be attenuated due to elastic and inelastic interactions with the produced medium~\cite{Gyulassy:1990ye,Wang:1992bz,Wiedemann:2009sh}. This phenomenon is also known as ``jet quenching'', originally proposed in~\cite{Bjorken:1982tu}, and is indirectly confirmed by measurements of spectra and correlations of high transverse momenta (\pt) hadrons at RHIC \cite{Adcox:2004mh,Adams:2005dq,Back:2004je,Arsene:2004fa} and LHC ~\cite{CMS:2012aa,Aamodt:2010jd,Aad:2015wga}. In these measurements, jet quenching is observed to have a dependence on event multiplicity and hadron \pt, and has provided significant insights, including the color opaqueness of the QGP. However, these findings are limited by intrinsic biases. For example, the leading hadron measurements are preferentially from the population of jets that have the least interaction with the medium. These measurements are also not sufficient to discriminate quantitatively between partonic energy loss formalisms or to extract key parameters such as the transport coefficient of the hot medium to precisely measure the stopping-power of the QGP (see Refs.~\cite{CasalderreySolana:2007zz,d'Enterria:2009am} for reviews). As jet quenching is intrinsically a partonic process, studies using hadronic observables blur essential physics due to the complexity of the theoretical description of hadronization and the sensitivity to non-perturbative effects. The measurement of jet structure and its modification in terms of energy flow rather than hadronic distributions promises a much closer connection to the underlying theory. Therefore a quantitative picture of jet quenching with respect to theoretical assumptions can be obtained through a full reconstruction of underlying parton kinematics, \ie, jet reconstruction~\cite{bass,Burke:2013yra}.

Complementary and robust jet measurements in heavy ion collisions became feasible with the beginning of the LHC heavy ion program. For example, measurements showed that the \pt of back-to-back dijet pairs becomes increasingly unbalanced as the centrality of the event increases (smaller impact parameters)~\cite{Aad:2010bu,Chatrchyan:2011sx,Chatrchyan:2012nia}. In these collisions jet pairs are also observed to be undeflected, \ie, their azimuthal angular correlations are independent of the collision centrality. Furthermore, measurements of jet shape, fragmentation functions, jet-track correlations, and missing \pt find that a significant fraction of the ``lost'' jet energy is observed to be
radiated via low-\pt particles far outside the jet cone~\cite{Chatrchyan:2011sx,Chatrchyan:2013kwa,Chatrchyan:2014ava,Khachatryan:2015lha,Khachatryan:2016erx}. The comparison of inclusive jets in heavy ion collisions with those in pp collisions can differentiate between competing models of parton energy loss mechanisms~\cite{majumder,Casalderrey-Solana:2015bww,Chien:2015hda}. Initial measurements of jet yields in central heavy ion collisions were compared to a pp baseline, and they are found to have a weak dependence on the jet \pt, with the low \pt region suffering slightly larger modification compared to the high \pt region~\cite{Aad:2014bxa,Adam:2015ewa}. However the interpretation of the jet modification results in nucleus-nucleus collisions and the understanding of their relation to the properties of the QGP requires detailed knowledge of all nuclear effects that could influence the comparisons with the pp system. The shape of the jet spectrum in proton-lead collisions is similar to that observed in pp collisions~\cite{Khachatryan:2016xdg,Adam:2015hoa,ATLAS:2014cpa}. This suggests the modification of the jet spectra observed in PbPb collisions is indeed an effect of the hot medium produced in these collisions.

For this analysis, the jet measurements are performed as a function of three experimental observables: the jet reconstruction distance parameter~\cite{Cacciari:2011ma}, the jet \pt, and the event centrality (related to the impact parameter of the incoming nuclei) of the collisions. The reference pp jet cross section is also measured and is compared to perturbative quantum chromodynamic (pQCD) calculations.
The observable of interest is the jet nuclear modification factor ($\RAA$), defined as,
\begin{equation}
\label{eq:raa}
\RAA= \frac{ \rd^{2}N^\mathrm{AA}_\text{jets}/\rd\pt\,\rd\eta} {\left<\TAA\right > \rd^{2}\sigma^{\pp}_\text{jets} / \rd\pt\,\rd\eta},
\end{equation}
where $N^\mathrm{AA}_\text{jets}$ is the jet spectrum measured in PbPb, $\sigma^{\pp}_\text{jets}$ is the jet cross section from pp collisions, and $\left<\TAA\right>$ is the nuclear overlap function averaged over the event class studied.
The quantity $\left<\TAA\right>$ is related to the mean number of nucleon-nucleon (NN) collisions $\left<\ncoll\right>$, and $\sigma^{\mathrm{NN}}_\text{inel}$, the nucleon-nucleon inelastic cross section, through $\left<\ncoll\right> = \left<\TAA\right> \, \sigma^\mathrm{NN}_\text{inel}$, and is calculated with a Monte Carlo Glauber model description of the nuclear collision geometry (for a review
see Ref.~\cite{Miller:2007ri}).

\section{The CMS detector and event selection}

\label{sec:event_selection}
The central feature of the CMS apparatus is a superconducting solenoid providing a magnetic field of 3.8\unit{T}. Charged-particle trajectories are measured with the silicon tracker that allows a transverse impact parameter resolution of ${\sim}15\mum$ and a \pt resolution of ${\sim}1.5$\% for particles with $\pt = 100$\GeVc. A PbWO$_{4}$ crystal electromagnetic calorimeter (ECAL) and a brass and scintillator hadron calorimeter (HCAL) surround the tracking volume. The forward regions
are instrumented with iron and quartz-fiber hadron forward calorimeters (HF).
A set of beam scintillator counters (BSC), used for triggering and beam halo rejection, is mounted on the inner side of the HF calorimeters. The very forward angles are covered at both ends
by zero-degree calorimeters (ZDC). A more detailed description of
the CMS detector, together with a definition of the coordinate system used and the relevant kinematic variables, can be found in Ref.~\cite{Chatrchyan:2008aa}.

The first level of the CMS trigger system, composed of custom hardware processors, uses information from the calorimeters to select the most interesting events in a fixed time interval of less than 4\mus. The high-level trigger (HLT) processor farm further decreases the event rate, from around 100\unit{kHz} to less than 1\unit{kHz}, before data storage.
The PbPb analysis uses minimum bias triggered and single-jet HLT data sets. The minimum bias events are characterized by the coincidence of signals in the two HF detectors or the forward and backward BSCs. The triggers used in the analysis are constructed from ECAL and HCAL energies requiring a single jet with $\pt > 55$, 65, and 80\GeVc. For pp collisions, the triggers require at least one jet with $\pt > 40$, 60, and 80\GeVc. The objects used in the HLT are jets reconstructed using the iterative-cone algorithm~\cite{CMS-PAS-JME-07-003} with distance parameter $\RR=0.5$.
The soft background in PbPb collisions is removed with the iterative pileup subtraction technique~\cite{Kodolova:2007hd}.
In order to extend the reach of the jet spectra, data sets from the high-\pt single-jet
triggers are combined together in both pp and PbPb. To reach lower jet~\pt in the PbPb data set, the minimum bias triggered events are added.

This analysis uses 166 \mubinv of PbPb collisions at $\rootsNN =2.76$\TeV recorded by CMS during the 2011 heavy ion run, as
well as 5.43\pbinv of pp collisions at the same collision energy recorded in
early 2013.
The event selection techniques developed for Ref.~\cite{Chatrchyan:2014ava} are employed. These include the identification of a primary vertex and the removal of contamination from beam background, ultra-peripheral and HCAL noise events.  The primary reconstructed vertex of selected events in the $z$ direction (beam axis) is constrained to be within ${\pm}15$\unit{cm} of the center of the detector. After these selections, events with more than one PbPb collision occurring in the same beam crossing remain and are later referred to as pileup.
Utilizing the sensitivity of the ZDC to spectator nucleons and of the HF to particles produced in the collisions, these pileup events  $(0.2\%)$ are removed by comparing the energy deposited in the ZDC to the HF.
This is further substantiated by counting the number of fully reconstructed jets with $\pt>50\GeVc$ and comparing this to the number of tracker pixel hit counts, since pileup events tend to have large pixel counts for the same number of jets. The selection for pileup events in data does not remove any events from the simulation. This procedure was checked by individually studying a representative sample of the rejected events.

Simulated dijet events are generated using \PYTHIA 6.4.23 Tune Z2~\cite{bib_pythia} for pp collisions at 2.76\TeV center-of-mass energy. For comparison to PbPb data, these \PYTHIA events are embedded into a simulated PbPb event, generated by {\textsc{hydjet}} (version 1.8)~\cite{Lokhtin:2005px}. The {\textsc{hydjet}} simulations are generated with jet quenching enabled in order to match the distribution of high-\pt jets in a minimum-bias data set. The {\textsc{hydjet}} simulations are tuned to represent a minimum bias background measured in CMS collisions of PbPb at $\rootsNN = 2.76\TeV$.
Collision centrality is classified with the standard CMS heavy ion technique~\cite{Chatrchyan:2014ava} using the total sum of the transverse energy in the HF towers, divided in percentiles according to the minimum bias samples. This distribution is divided into centrality bins, each representing 0.5\% of the total
nucleus-nucleus interaction cross section. For this analysis, the results are collected in six bins corresponding to the most central (\ie smallest impact parameter) 5\% of
the events, denoted 0\%--5\%, as well as bins of 5\%--10\%, 10\%--30\%,
30\%--50\%, 50\%--70\% and 70\%--90\%. The centrality of an event can be correlated with the impact parameter, as well as with  $\left<\npart\right>$, the average number of nucleons in the nuclei that participate in the collision, using MC Glauber model
calculations~\cite{Miller:2007ri}.

\section{Jet reconstruction and selection}
\label{sec:jet_reco}

Similar to Refs.~\cite{Chatrchyan:2012gt,Chatrchyan:2012nia,Chatrchyan:2014ava,Chatrchyan:2011sx},
jet reconstruction in heavy ion collisions in CMS is performed with the sequential \ak clustering
algorithm via the \FASTJET framework~\cite{Cacciari:2011ma}.
The jet clustering is performed using particle-flow (PF)~\cite{CMS:2009nxa,CMS:2010byl} candidates that combine information from the individual CMS detector systems. Different particle types (charged and neutral hadrons, electrons, muons, and photons) are reconstructed. The \ak distance parameters used are $\RR =$ 0.2, 0.3, and 0.4.

For PbPb collisions, the soft underlying event (background) is removed from the jets with an iterative subtraction technique described in Ref.~\cite{Kodolova:2007hd}. In this procedure, the PF candidates are grouped in towers that correspond to the calorimeter geometry. Jets are selected with $\abs{\eta}<2$ to ensure that they are fully contained within the CMS tracker up to a distance parameter of 0.4. Detector-based $\eta$ and \pt dependent energy correction factors~\cite{Chatrchyan:2011ds} are applied to the jets. The raw jet \pt of a jet is the \pt before any of the detector-based corrections are applied. To study the background in PbPb events, data and \PYTHJET simulations are compared. The correction to the jet \pt obtained from this iterative subtraction technique (called ``raw subtracted \pt{}"), for a jet with distance parameter $R^\text{jet}$ is estimated by taking the difference between the sum of all the PF candidate \pt in a $\Delta \RR < R^\text{jet}$ cone and the raw jet \pt. The $\Delta \RR$ is defined as the distance of the PF candidate from the reconstructed jet axis in the $\eta$-$\phi$ plane:
\begin{equation}
\label{eq:DeltaR}
\Delta \RR=\sqrt{(\Delta\phi_\text{candidate, jet})^2+(\Delta \eta_\text{candidate, jet})^2}.
\end{equation}

\begin{figure}[htb]
  \centering
    \includegraphics[width=\cmsFigWidth]{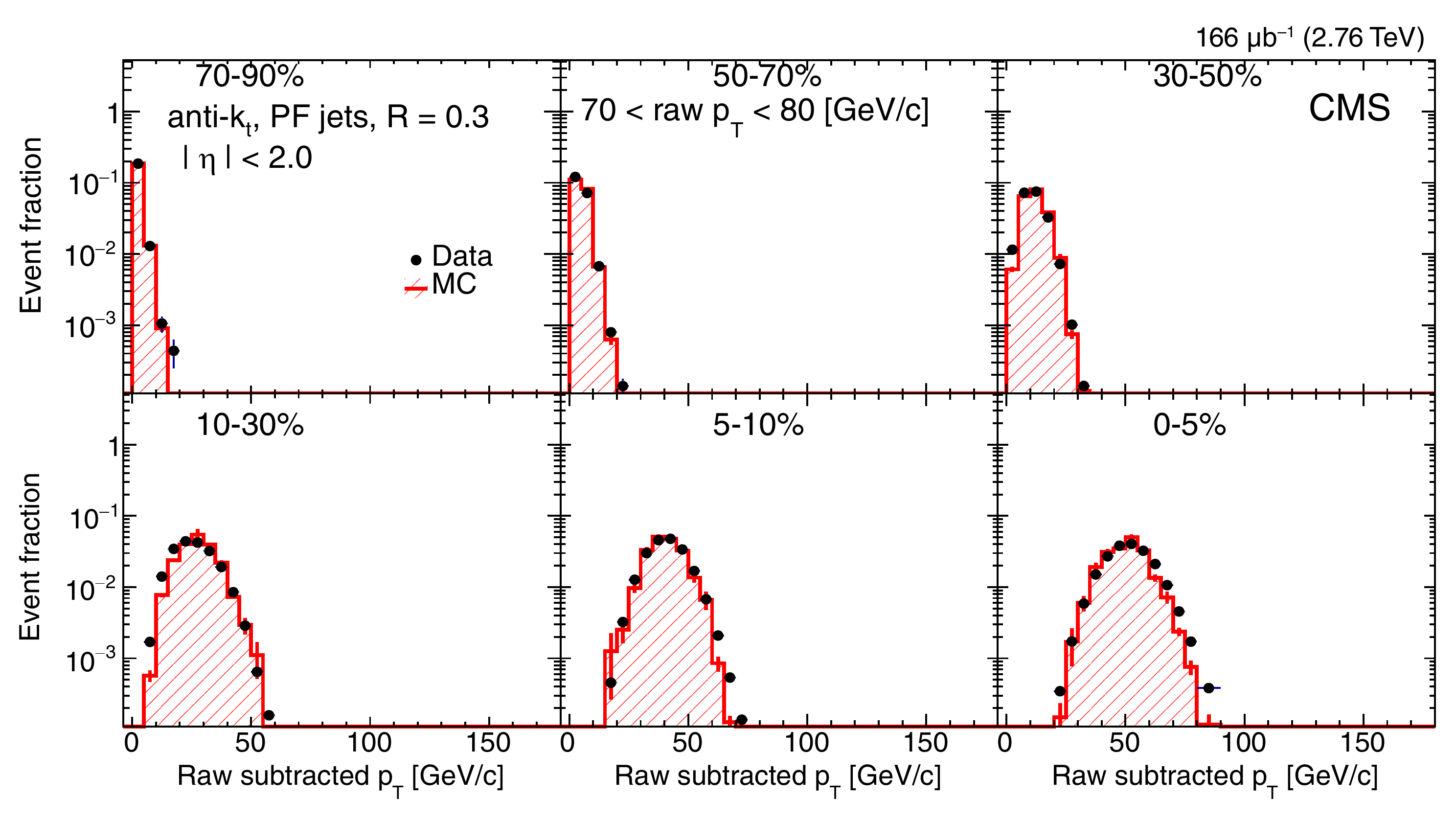}
    \includegraphics[width=\cmsFigWidth]{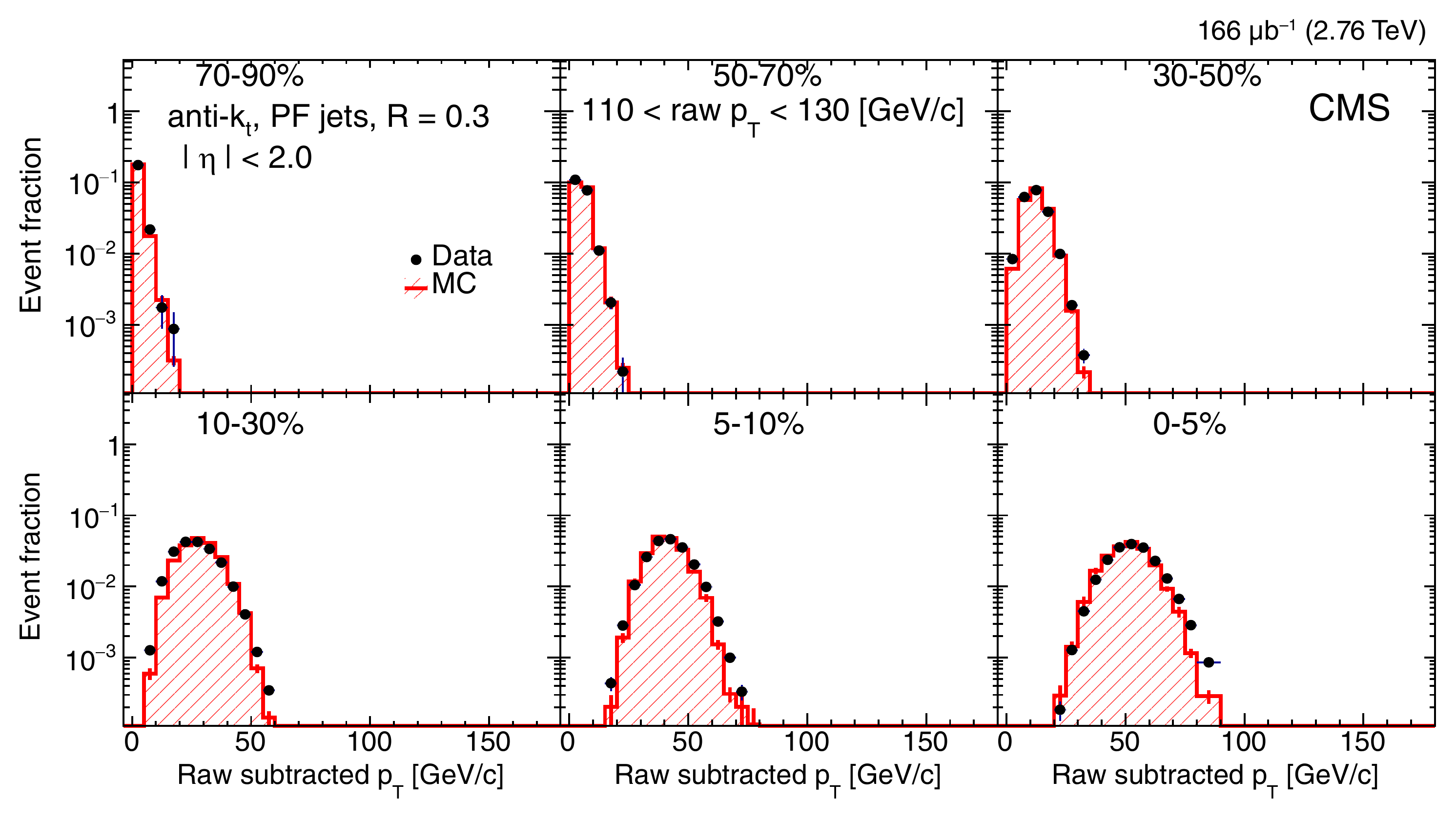}
    \caption{
      Raw subtracted \pt for jets reconstructed with the \ak algorithm and a distance parameter of $\RR = 0.3$, in the ranges $70<  \text{jet \pt} < 80$\,[\GeVcns{}] (top panels) and $110< \text{jet \pt} < 130$\,[\GeVcns{}] (bottom panels). This quantity is found by taking the difference of the sum of PF candidates within the jet cone and raw jet \pt. Solid symbols show data, and the histogram is from \PYTHJET generated events.
    }
    \label{fig:AvbkgdpT_wjid_R3}

\end{figure}

The distributions of raw subtracted \pt for $\RR = 0.3$ jets, from peripheral to central collisions are shown in Fig.~\ref{fig:AvbkgdpT_wjid_R3} for two different reconstructed jet \pt selections. Data are shown with filled circles and simulations with histograms. There is a good agreement between the two in all centralities and jet \pt bins. A similar level of agreement is also seen for $\RR=0.2$ and $\RR=0.4$.

\begin{figure}[htb]
  \centering
    \includegraphics[width=0.49\textwidth]{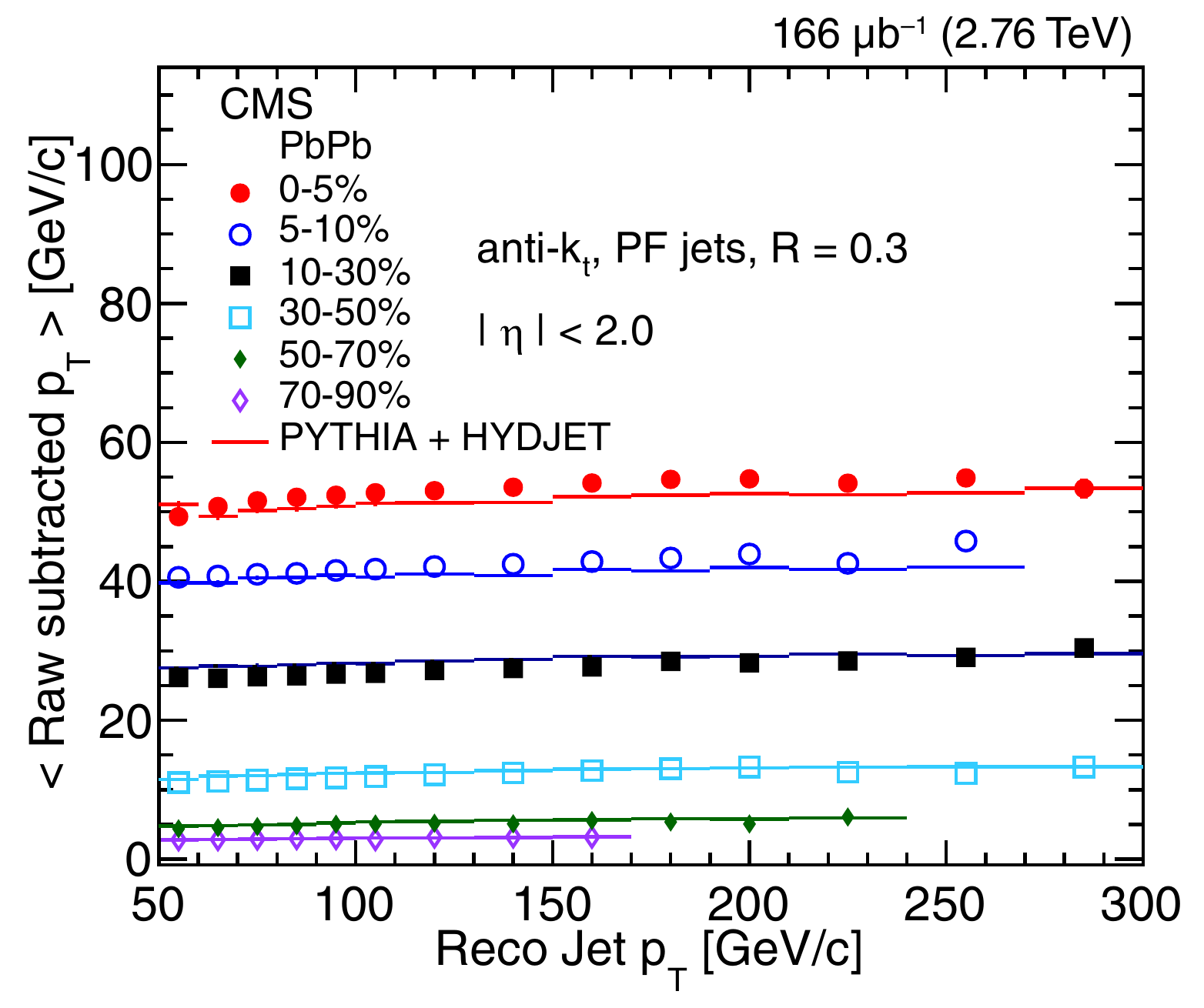}
     \includegraphics[width=0.49\textwidth]{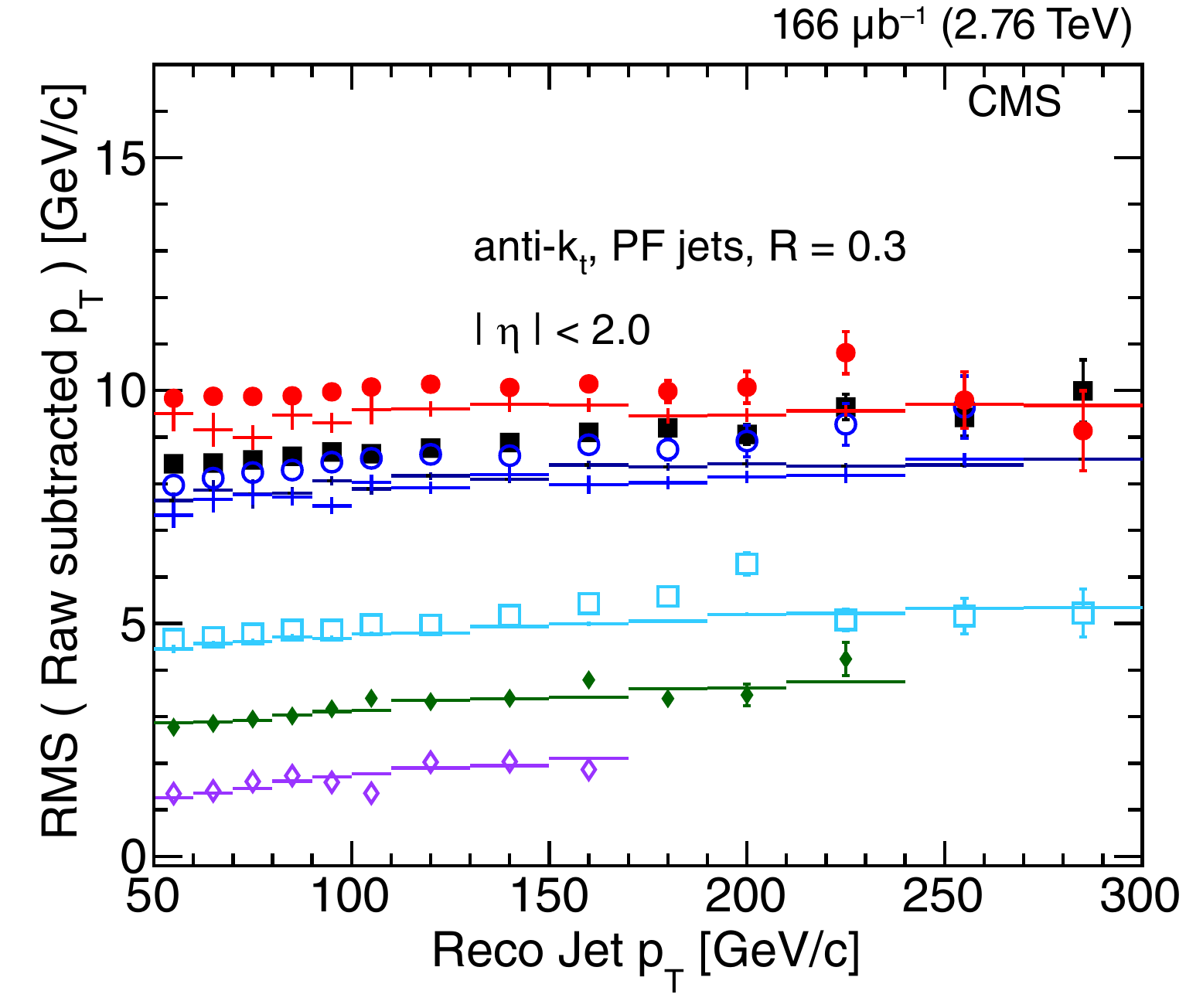}
    \caption{
      Average raw subtracted \pt (\cmsLeft) and its RMS (\cmsRight) for PF jets reconstructed with the \ak algorithm, with a distance parameter $\RR = 0.3$. Symbols represent data, and lines show \PYTHJET simulated events.
    }
    \label{fig:Bkgd_Mean_RMS_R3}

\end{figure}

The average raw subtracted \pt and its root mean square (RMS) values are shown in Fig.~\ref{fig:Bkgd_Mean_RMS_R3} as a function of the reconstructed jet \pt, from central to the most peripheral collisions. Data are shown with markers and are compared with the \PYTHJET generated events shown as histograms. The average raw subtracted \pt decreases, from the most central to peripheral events, as expected, and distributions show reasonable agreement between data and \PYTHJET.

\subsection{Data driven correction}
\label{sec:backgroundjet}
Although the soft background is primarily removed with the iterative-pileup subtraction, fluctuations in this background can result in misreconstructed jets that do not originate from hard scattering.
A method to remove this contamination, used in other experiments~\cite{Aad:2014bxa,Adam:2015ewa}, is to select jets with a requirement on the leading charged-particle track or calorimeter energy deposit among the constituents of the jet.
However, this method can bias to preferentially select jets with hard fragmentation, distorting the low-\pt region. In CMS, tracks are reconstructed with a minimum \pt of $0.15\GeVc$, thus removing any such potential bias.

In this analysis, a novel data-driven technique, based on control regions in data, is introduced to derive the spectrum of misreconstructed jets from the minimum bias sample. This spectrum is then subtracted from the jet-triggered sample. Two methods, operating in different kinematic regimes, are combined to get a correction factor. The first method (labeled the trigger object method) selects all events with a leading HLT jet \pt of less than $60\GeVc$ as
a control sample potentially containing misreconstructed jets. This \pt threshold is chosen based on analysis of random cones in minimum bias events, with the leading and subleading jets removed.
The second method (labeled the dijet method), performed in parallel with the first method, selects minimum bias events with dijets, which can originate either from a hard scattering or fluctuating background. There are two thresholds defined in this method, one for the leading jet ($\pt^{\text{min} 1}$) and another for the subleading jet ($\pt^{\text{min} 2}$) in the reconstructed event. If an event fails any of the following selections, it is tagged as a background event. An event is tagged as a signal if it passes all of the criteria: Leading jet $\pt > \pt^{\text{min} 1} $ and $\Delta \phi_{j1, j2} > 2\pi/3$ and subleading jet $\pt > \pt^{\text{min} 2}$.
To choose the thresholds for the dijet selection, the mean and RMS of the subtraction step in the iterative subtraction algorithm are mimicked by applying a cutoff on the transverse energies of the PF towers used in the random cone study. The RMS of the background subtracted event energy distribution is used as an estimate of the fluctuation. The thresholds are set as follows: $ \pt^{\text{min} 1} = 3\, \mathrm{RMS}$ for the leading jet, and $ \pt^{\text{min} 2} = 1.8\, \mathrm{RMS}$ for the subleading jet, to allow for jet modification in the medium.

\begin{figure*}[htb]
   \centering
   \includegraphics[width=0.98\textwidth]{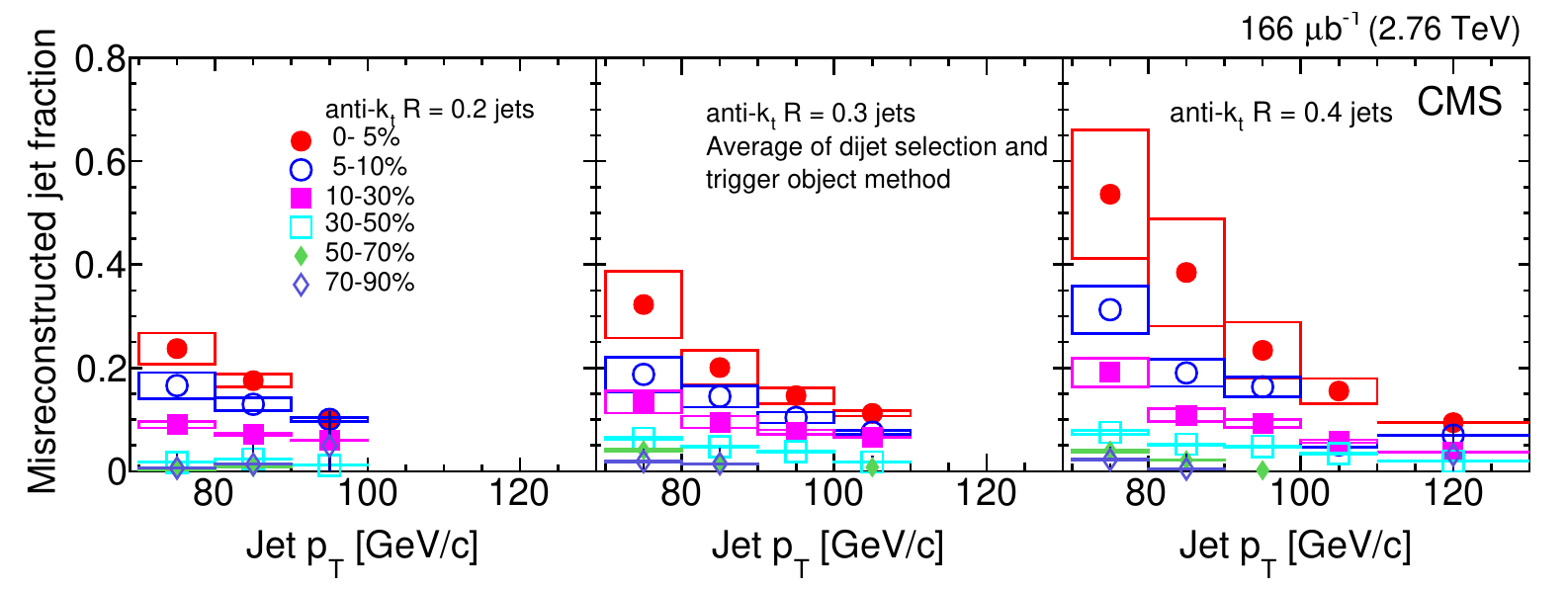}
   \caption{Misreconstructed jet fraction of the inclusive jet spectra, derived from the minimum bias sample, as a function of reconstructed jet \pt, for various centralities and three different distance parameters (left: $\RR = 0.2$, center: $\RR = 0.3$, and right: $\RR = 0.4$). The correction factor is the average of the dijet selection and trigger object methods discussed in the text. }
   \label{fig:minbias_backgroundrate_R3}
\end{figure*}

Since these two methods operate in different kinematic regimes, the average of the two is used to estimate the data driven correction factor for misreconstructed jet rates as can be seen in Fig.~\ref{fig:minbias_backgroundrate_R3}, as a function of the jet \pt. These rates for different distance parameters are shown in the different panels (left: $\RR = 0.2$, center: $\RR = 0.3$, and right: $\RR = 0.4$). The symbols correspond to the centrality bins in the analysis.
The minimum bias background jet spectra are then normalized to a per-event yield and the background is removed from the measured jet spectra, resulting in an inclusive jet spectrum without fragmentation bias. The correction, estimated in a similar way from {\PYTHIA} dijet events, where one does not expect any background, is added as an additional systematic uncertainty, ranging from 6\% at 70 GeV to 1\% at 100 GeV. The data driven method was also applied to {\textsc{pythia+hydjet}} simulations without quenching and, using the same \pt threshold, this yielded a recovery efficiency of greater than 98\% for signal jets, which is well within systematic uncertainties as described in Sec.~\ref{sec:sys}.

\subsection{Unfolding studies}
An unfolding method is required to remove the smearing and bin migration in jet \pt due to detector resolution, and to extract the jet cross section measurement. Three different techniques are used to determine the final jet \pt
spectra: Single value decomposition (SVD), Bayesian, and a bin-by-bin unfolding
technique~\cite{Elden:1982wp,D'Agostini:1994zf,Hansen:1998rd,Adye:2011gm}. Results
presented here are based on the SVD technique, while the others are used as a cross-check, giving consistent results within their respective uncertainties.
The three aforementioned procedures use a response matrix from \PYTHJET
of reconstructed jets, matched to generator-level jets in the $\eta$-$\phi$ space, that originate
from the \PYTHIA QCD hard scattering.

The SVD unfolding is performed with a
regularization parameter, which is optimized for each
centrality class and each jet resolution. The simulation and data used in
unfolding have a reconstructed jet \pt larger than 50\GeVc for all distance parameters,
with unfolded results reported for jets larger than 70\GeVc.

\section{Systematic uncertainties}
\label{sec:sys}

The systematic uncertainty is calculated from a number of
sources and is shown in Table ~\ref{table:systematics}. For $\RR = 0.3$ jets, in the low $\pt <80$\GeVc region, a large contribution to the jet yield uncertainty in PbPb collisions is from the data driven corrections (20\%). The data driven systematic uncertainty is estimated from the overlap of the two different methods (trigger object and dijet methods as described in Sec.~\ref{sec:backgroundjet}) along with an additional uncertainty of 1-6\% across all jet $\pt$, centrality ranges, and jet distance parameters determined from its application on a {\textsc{pythia}} sample. The jet energy scale (JES) uncertainty ranges from 6--32\% (from peripheral to central events), varying due to the uncertainty in the heavy ion tracking and the quark/gluon fragmentation. The fragmentation difference is included in the JES uncertainty for pp, but is extended for PbPb jets due to expected asymmetric jet quenching effects for quark and gluon jets. The jet response matrix is smeared by 1\%, at both the generator and reconstructed levels to account for variations in the simulations. Separately the regularization parameter used for the unfolding is varied between 4 and 8 resulting in at most 8\% systematic uncertainty for the PbPb jet yield and at most 2\% for the pp jet cross section.

A residual jet energy correction, using the dijet balance method~\cite{Chatrchyan:2011ds}, is derived and applied to the jets from pp collisions. It corresponds to less than 1\% correction to the jet \pt. The jet energy resolution (JER) uncertainty is
estimated for each \pt bin in the analysis and is found to be at most
3\%, for both pp and PbPb.
Studies of the underlying event fluctuations in jet-triggered and minimum bias events show a contribution of up to 5\% to the uncertainty of reconstructed jet yields based on differences between data and \PYTHJET quantified in the right side of Fig.~\ref{fig:Bkgd_Mean_RMS_R3}.
The contributions due to jet reconstruction efficiency, detector noise, and unfolding response matrix smearing are about 1\% each.

Since in PbPb, the per-event jet yield is being measured, there is a 3\% uncertainty on the number of minimum bias events and there is no uncertainty quoted for the luminosity. For the pp cross section, there is a 3.7\% uncertainty in the integrated luminosity~\cite{CMS-PAS-LUM-13-002}.
Systematic uncertainties, from different contributions to the jet
$\RAA$, are summed in quadrature with an overall uncertainty of 19--40\%, from
peripheral to central collisions for $\RR  =  0.3$\,jets. Detailed systematic uncertainties for different $\RR$ and two representative jet \pt ranges are shown in Table~\ref{table:systematics}.

\begin{table*}[htb]
  \centering
  \topcaption{Summary of the systematic uncertainties in the PbPb jet yield for the central (0--5\%), peripheral (70--90\%) bins, and the pp jet cross section. Each column showcases the total systematic uncertainties for the corresponding source for the different $\RR$ and two jet \pt ranges \ie $70 < \text{jet \pt}< 80\,[\GeVcns{}]$ and $250 < \text{jet \pt} < 300$\,[\GeVcns{}]). The $\TAA$ uncertainties are not shown in the table. Other sources mentioned in the text that are smaller than 1\% are not listed explicitly below.}
  \label{table:systematics}
  \cmsTable{\begin{scotch}{llrrr{c}@{\hspace*{5pt}}rrr}
    & \multirow{2}{*} {Source}
    & \multicolumn{3}{c} {$70 < \text{jet \pt}< 80$\,[\GeVcns{}]} && \multicolumn{3}{c} {$250 < \text{jet \pt}< 300$\,[\GeVcns{}] } \\ \cline{3-5}\cline{7-9}
    && $\RR = 0.2$ & $\RR = 0.3$ & $\RR = 0.4$ && $\RR = 0.2$ & $\RR = 0.3$ & $\RR = 0.4$ \\
    \hline
    \PbPb: & Data driven correction & 13\% & 20\% & 27\% && \NA & \NA & \NA \\
    (0-5\%)& JES \& unfolding & 32\% & 32\% & 48\% && 19\% & 19\% & 21\% \\
    & JER & 3\% & 3\% & 3\% && 3\% & 3\% & 3\% \\
    & Underlying event & 5\% & 5\% & 5\% && \NA & \NA & \NA \\
    \hline
    \PbPb: & Data driven correction & 8\% & 10\% & 12\% && \NA & \NA & \NA \\
    (70-90\%)& JES \& unfolding & 16\% & 16\% & 18\% && \NA & \NA & \NA \\
    & JER & 3\% & 3\% & 3\% && \NA & \NA & \NA \\
    & Underlying event & 5\% & 5\% & 5\% && \NA & \NA & \NA \\
    \hline
    \pp: & JES \& unfolding & 7\% & 7\% & 6\% && 5\% & 4\% & 5\% \\
    & JER & 3\% & 3\% & 3\% && 2\% & 2\% & 2\% \\
    & Integrated luminosity & 3.7\% & 3.7\% & 3.7\% && 3.7\% & 3.7\% & 3.7\% \\
  \end{scotch}}
\end{table*}

\section{Results}
\label{sec:results}

\begin{figure*}[htb]
  \centering
    \includegraphics[width=0.9\textwidth]{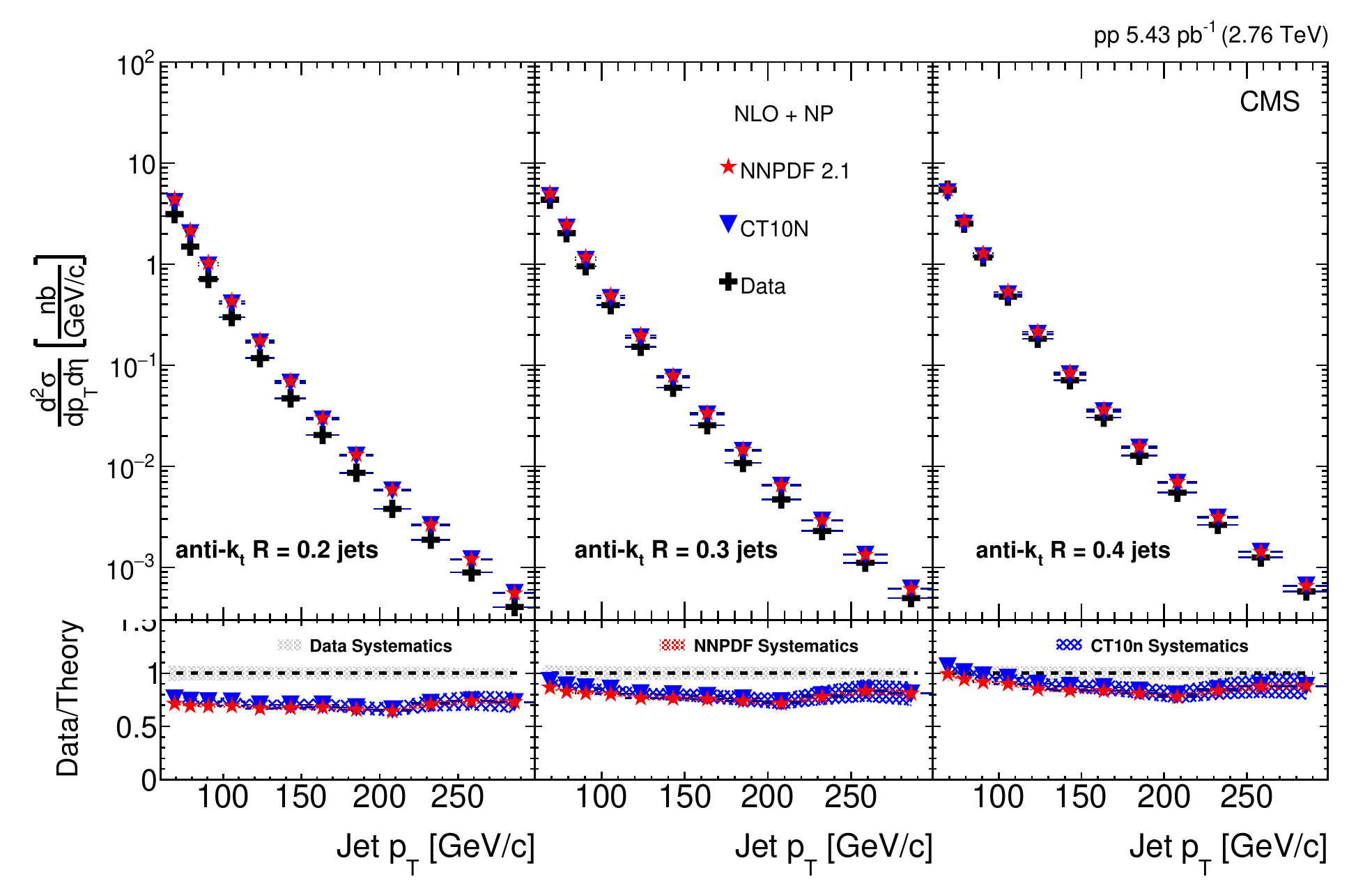}
    \caption{
Comparison of the inclusive jet cross section for \ak jets with distance parameters of $\RR  =  0.2$ (left), 0.3 (middle) and 0.4 (right), measured for pp collisions at 2.76\TeV (black plus markers), and NLO calculations, at the same collision energy, with NNPDF 2.1 (red star) and CT10N (blue triangle), with their respective NP corrections added. The bottom panels show the ratio of measured cross section to theory calculations. The systematic uncertainties for data are shown in the gray shaded band, while the systematic uncertainties in the NLO calculations are shown with the respective color shaded bands.
    }
    \label{fig:JetSpectraSVD_ppNLO}

\end{figure*}

The inclusive jet cross sections in pp collisions at 2.76\TeV are shown in Fig.~\ref{fig:JetSpectraSVD_ppNLO} for three different distance parameters. A comparison is made to next-to-leading-order (NLO)~\cite{Wobisch:2011ij} calculations of quantum chromodynamics.
These calculations are shown for two parton distribution functions (PDF) sets: NNPDF 2.1~\cite{Ball:2012cx} (red stars), and CT10N~\cite{Gao:2013xoa} (purple triangles) including non-perturbative (NP) contributions such as multi-parton interactions and hadronization.
Contributions to the jet cross section from NP effects are not inherently included in pQCD calculations due to a lower scale cutoff of a few \GeVc.
Thus, the NP correction factors need to be added and are computed as the ratio of cross sections calculated with
leading order (LO) + parton shower (PS) + multi-parton interactions + hadronization to
 LO+PS~\cite{Wobisch:2011ij}.
The bottom panel of Fig.~\ref{fig:JetSpectraSVD_ppNLO} shows the ratio of the data for jet cross sections in pp collisions to theoretical calculations, with the measured jet cross section from pp collisions for different distance parameters. The agreement with data gets better at larger distance parameters. In Ref.~\cite{Khachatryan:2015luy} the ratio tends closer to unity for jets with $\RR = 0.7$. The theoretical uncertainties shown are due to variations of the strong coupling constant and the parton shower, factorization scales involved in the NLO calculations for the different PDF sets.

\begin{figure}[htb]
  \centering
    \includegraphics[width=\cmsFigWidth]{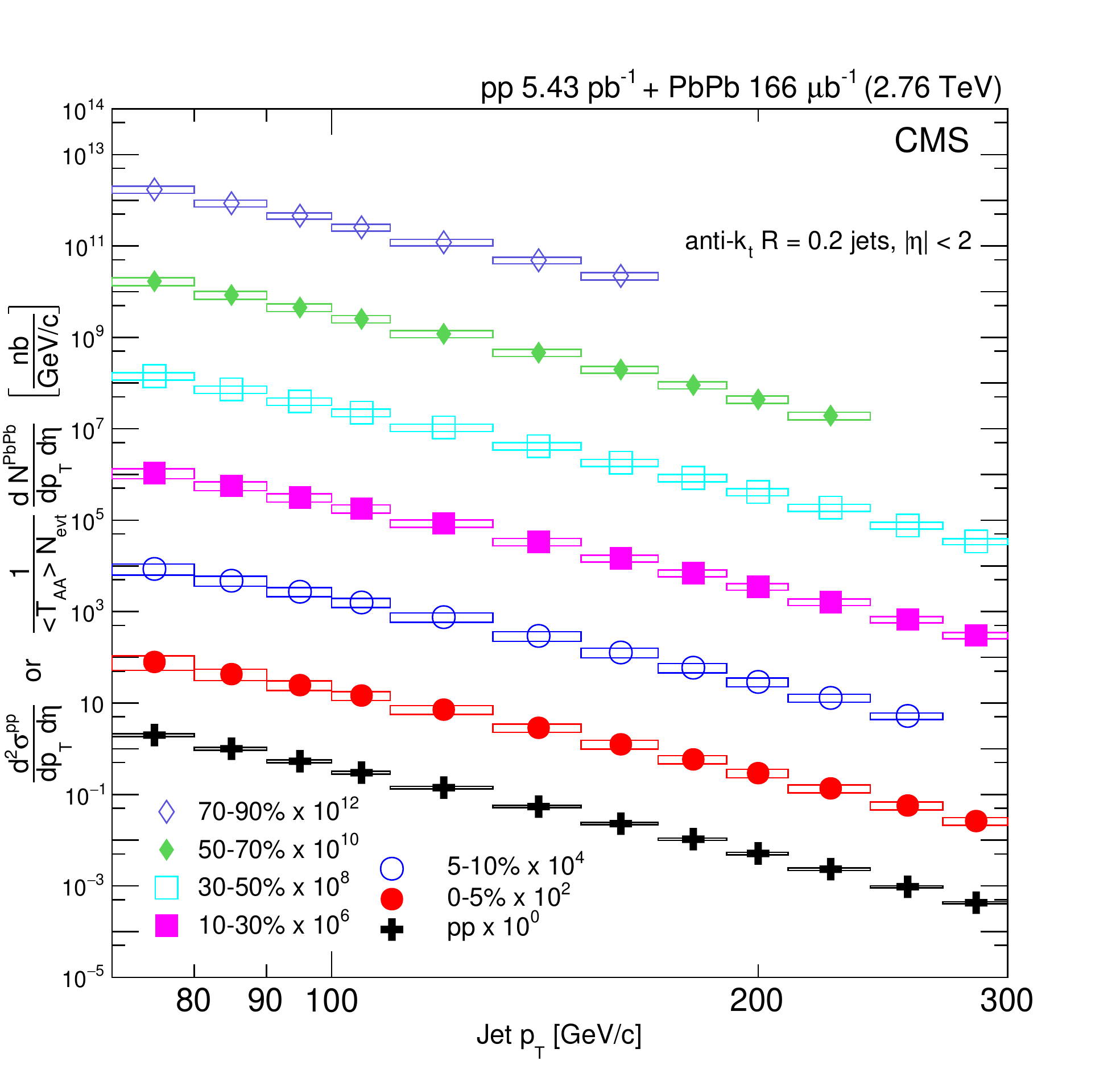}
    \caption{
      Inclusive jet spectra for PbPb jets of distance parameter $\RR  =  0.2$, in different centrality bins, and pp reference data. The PbPb jet spectra for different centrality classes are scaled by $\left<\TAA\right>$ and multiplied by a different factor for better visualization. Vertical bars represent statistical uncertainty (too small to see on this scale) with the systematical uncertainty in the colored boxes around the data points.
    }
    \label{fig:JetSpectraSVD_ak2}

\end{figure}

\begin{figure}[htb]
  \centering
    \includegraphics[width=\cmsFigWidth]{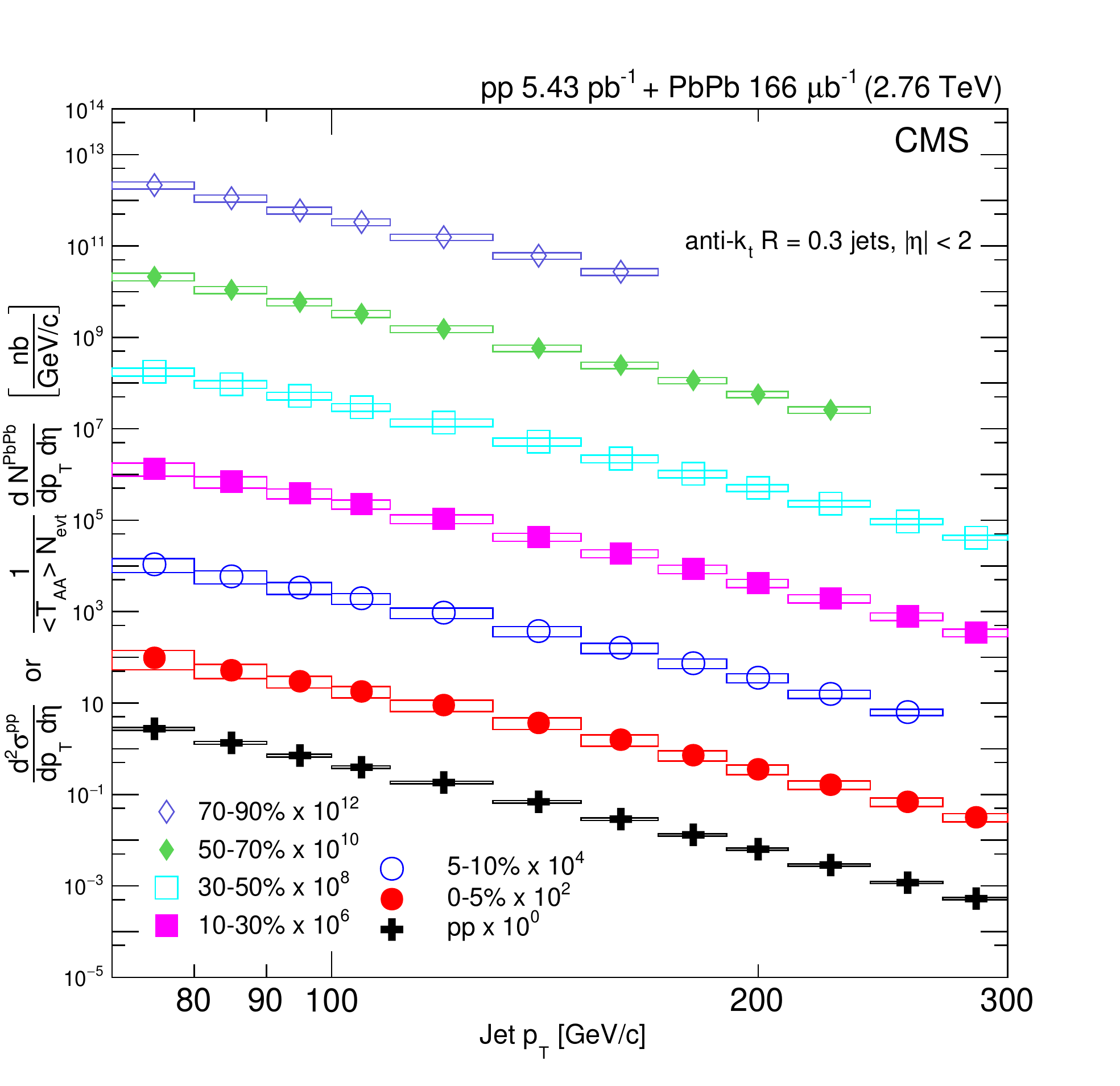}
    \caption{
      Inclusive jet spectra for PbPb jets of distance parameter $\RR  =  0.3$, in different centrality bins, and pp reference data. The PbPb jet spectra for different centrality classes are scaled by $\left<\TAA\right>$ and multiplied by a different factor for better visualization. Vertical bars represent statistical uncertainty (too small to see on this scale) with the systematical uncertainty in the colored boxes around the data points.    }
    \label{fig:JetSpectraSVD_ak3}

\end{figure}

\begin{figure}[htb]
  \centering
    \includegraphics[width=\cmsFigWidth]{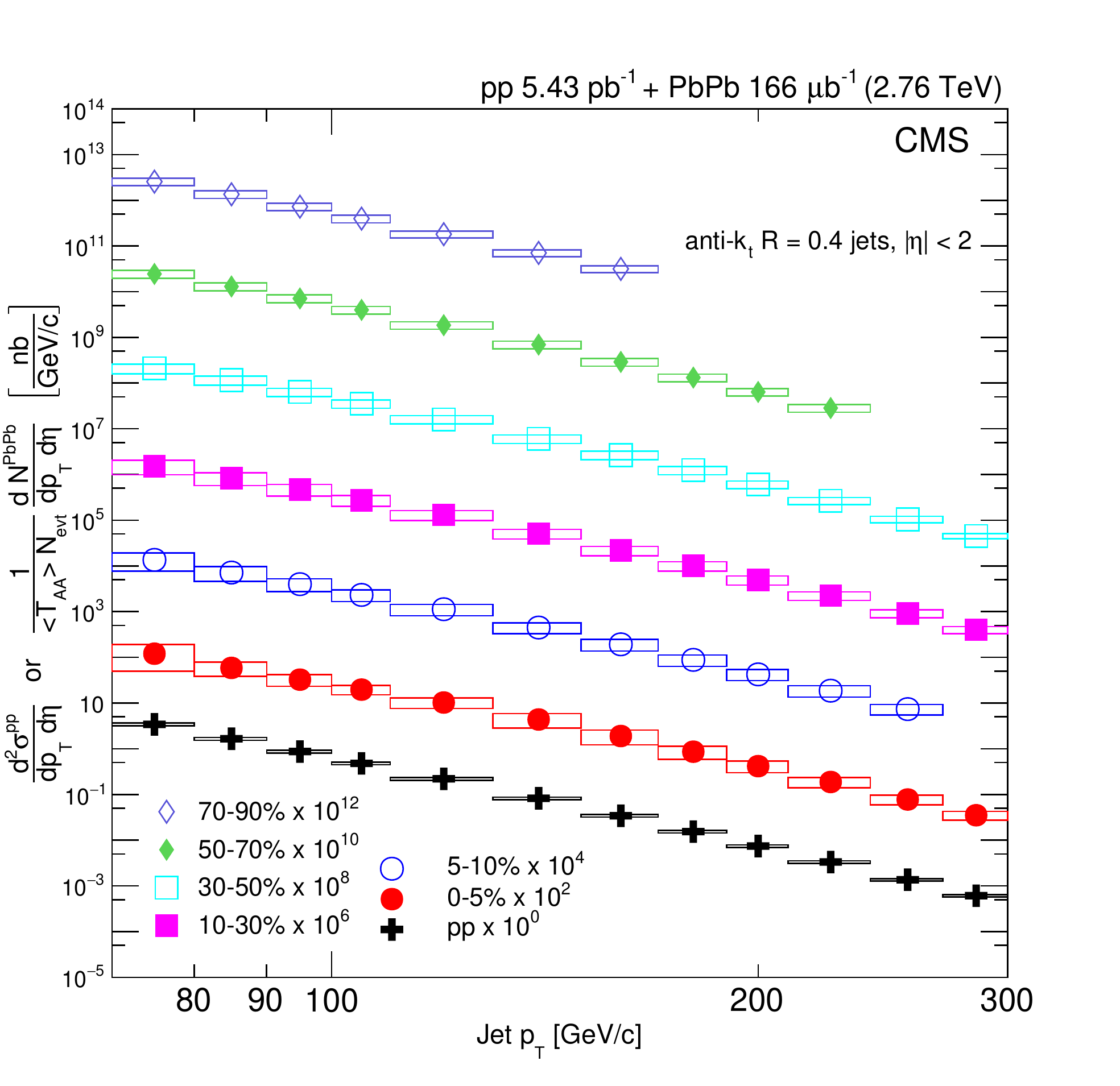}
    \caption{
      Inclusive jet spectra for PbPb jets of distance parameter $\RR  =  0.4$, in different centrality bins, and pp reference data. The PbPb jet spectra for different centrality classes are scaled by $\left<\TAA\right>$ and multiplied by a different factor for better visualization. Vertical bars represent statistical uncertainty (too small to see on this scale) with the systematical uncertainty in the colored boxes around the data points.    }
    \label{fig:JetSpectraSVD_ak4}

\end{figure}

The unfolded jet cross sections for PbPb and pp events are shown in
Figs.~\ref{fig:JetSpectraSVD_ak2}-\ref{fig:JetSpectraSVD_ak4} for
different distance parameters. The PbPb spectra are normalized by the
number of minimum bias events, and are scaled by
$\left<\TAA\right>$, with each centrality multiplied by a different factor, to
separate the spectra for better visualization. The pp reference data is normalized to the integrated luminosity of the analyzed data set.
The high \pt cutoffs for the spectra (hence also the $\RAA$) are dictated by statistical limitations.

\begin{figure*}[htb]
  \centering
    \includegraphics[width=0.98\textwidth]{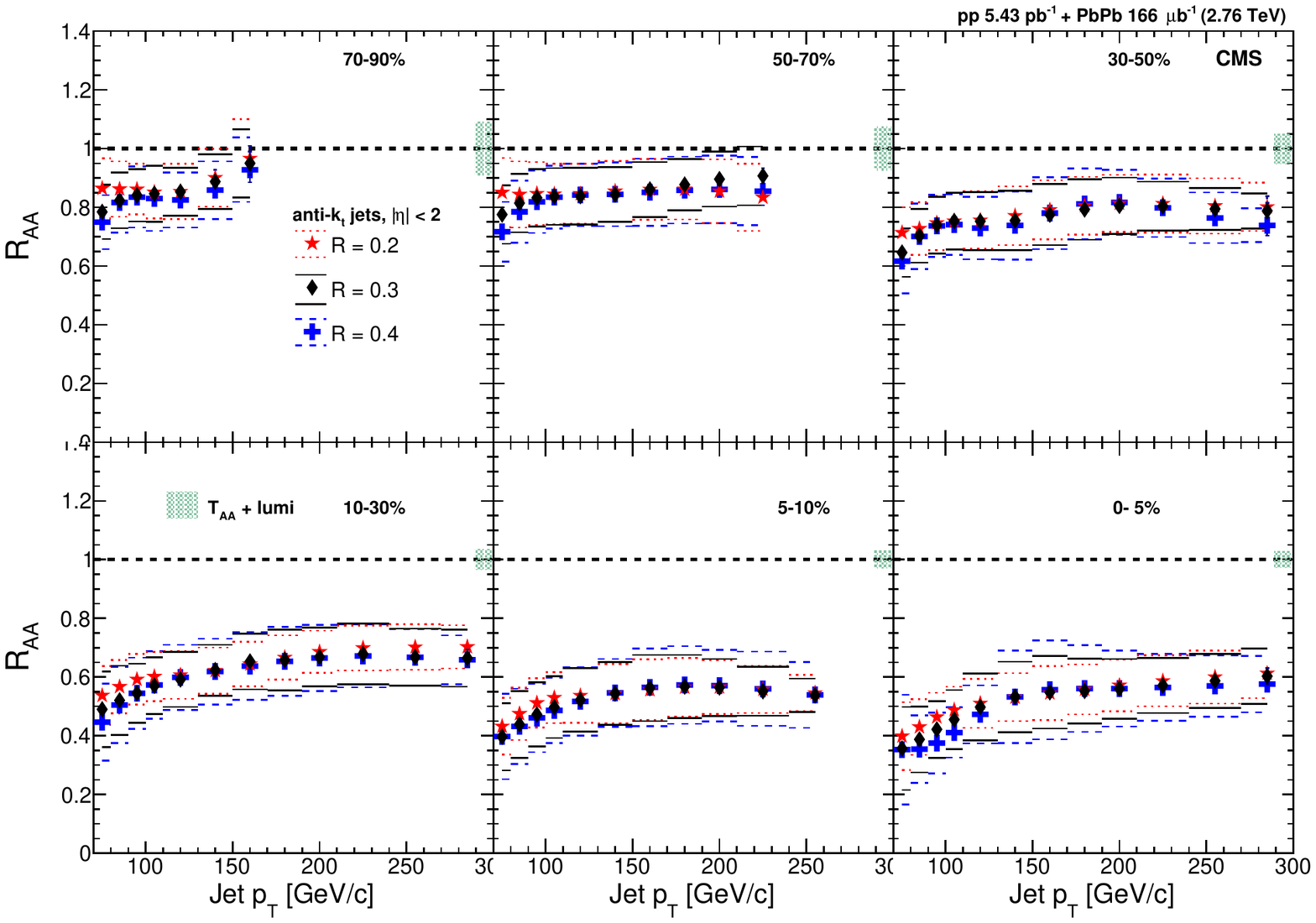}
    \caption{Inclusive jet \RAA as a function of the jet \pt, for \ak jets with distance parameters $\RR  =  0.2$ (red stars), 0.3 (black diamonds), and 0.4 (blue crosses) for different centrality bins. The vertical bars (smaller than the markers) indicate the statistical uncertainty and the systematic uncertainty is represented by the bounds of the dotted, solid, and dashed horizontal lines. The uncertainty boxes at unity represent the \TAA and luminosity uncertainty.}
    \label{fig:UnfoldRAA_akRadii}
\end{figure*}

The jet $\RAA$, found from the PbPb and pp spectra after all corrections including SVD unfolding, are shown for different distance parameters in Fig.~\ref{fig:UnfoldRAA_akRadii}. The jet \RAA decreases with increasing collision centrality in the range of the measured jet \pt. Within the systematic uncertainty, the jet \RAA shows the same level of suppression for the three distance parameters. Uncorrelated uncertainties remain too large to further elucidate the hierarchy of the jet distance parameter dependence of this \RAA measurement.

\begin{figure}[htb]
  \centering
    \includegraphics[width=0.48\textwidth]{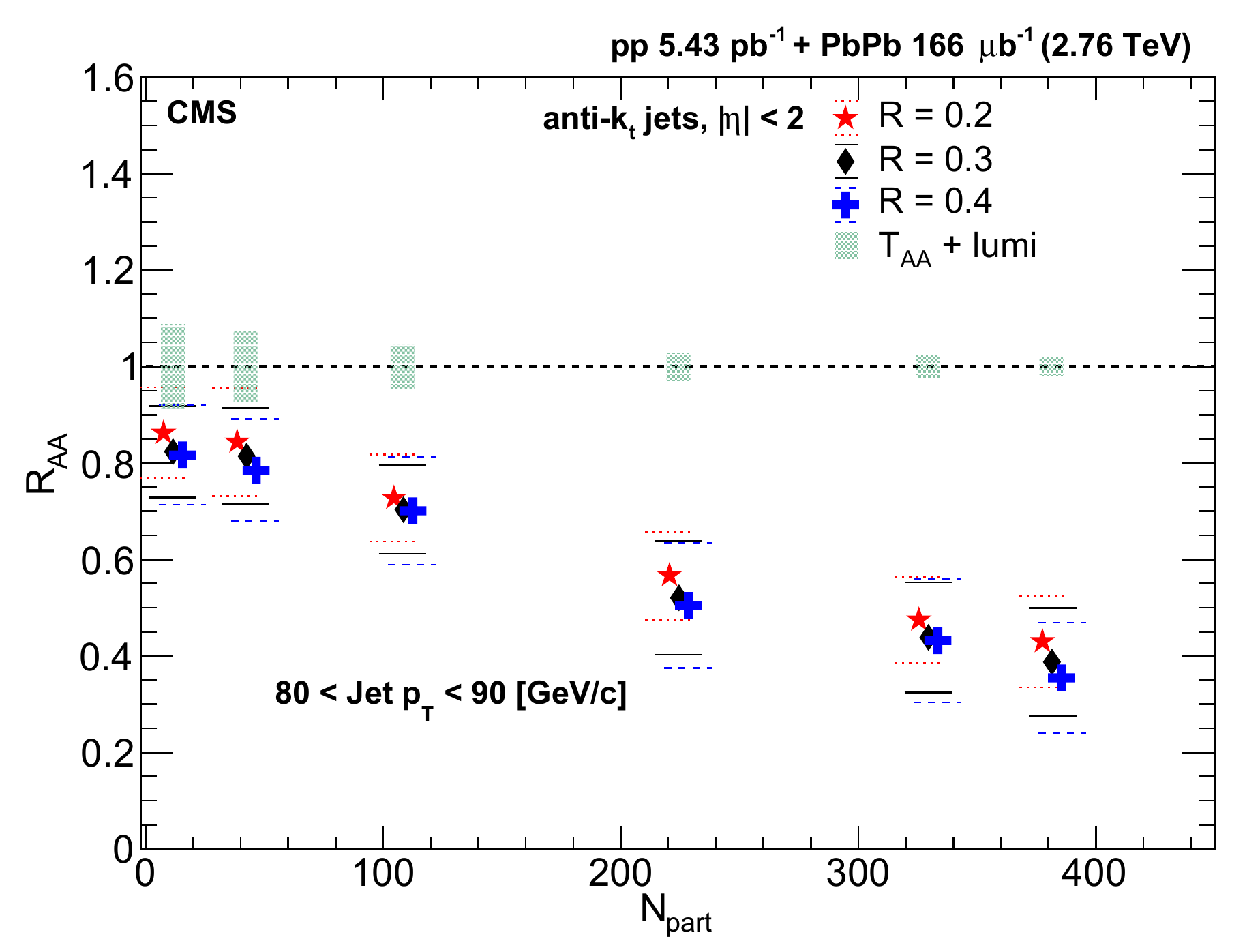}
    \includegraphics[width=0.48\textwidth]{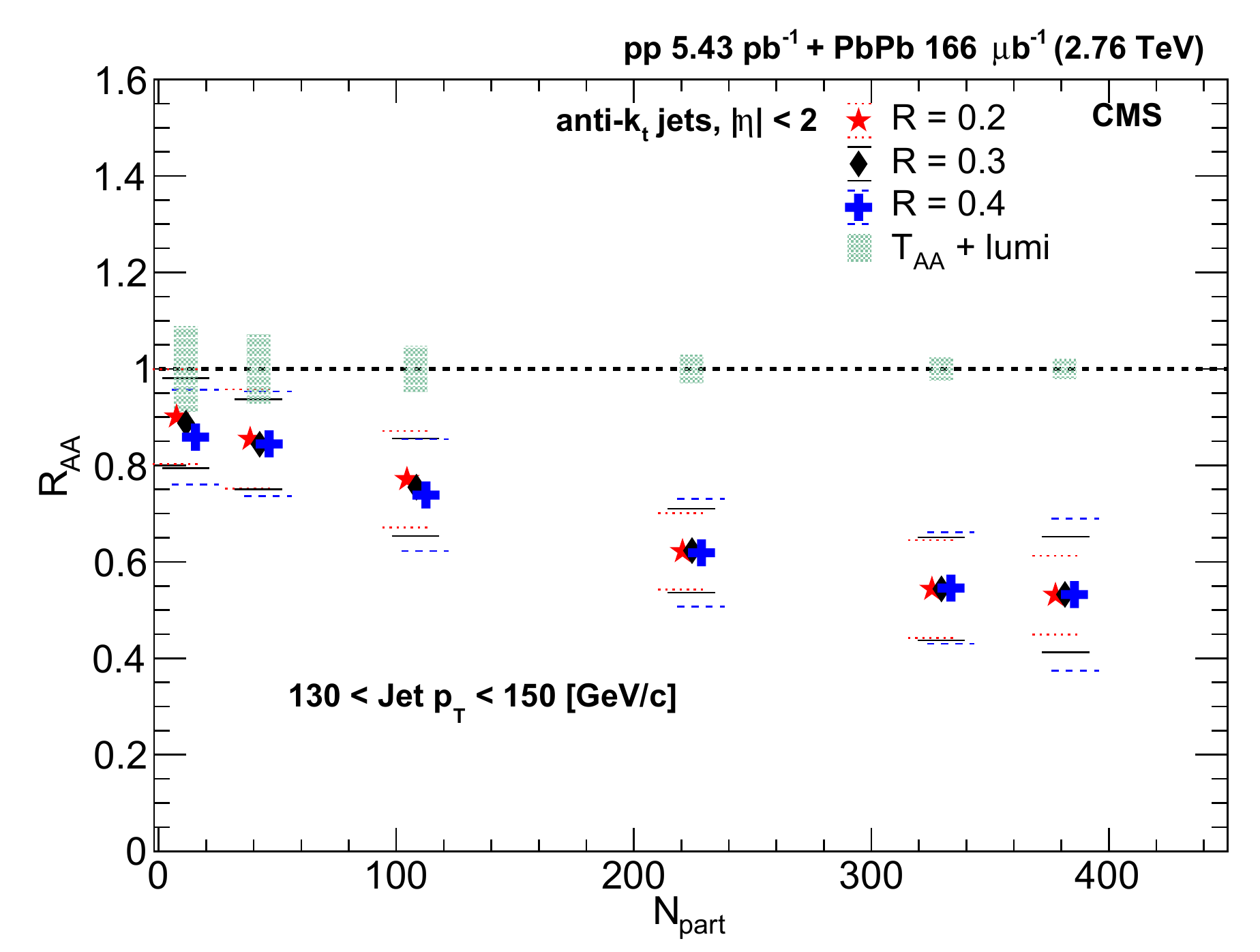}
    \caption{Inclusive jet \RAA for \ak jets with distance parameters $\RR  =  0.2$ (red stars), 0.3 (black diamonds), and 0.4 (blue crosses), as a function of the average $N_{\text{part}}$ for each collision centrality, for jets of $80 < \pt < 90$ and $130 < \pt < 150\,[\GeVcns{}]$, in the \cmsLeft and \cmsRight panels respectively. Points are shifted to the left ($\RR  = 0.2$) and right ($\RR  = 0.4$) for clarity. The statistical uncertainty is indicated by colored vertical lines (smaller than the markers). The systematic uncertainty is represented by the bounds of the dotted, solid, and dashed horizontal lines for the corresponding distance parameters. The uncertainty boxes at unity represent the \TAA and luminosity uncertainty.}
    \label{fig:UnfoldRAA_akRadii_Npart_20_eta_20}

\end{figure}

To focus on the centrality dependence of the jet $\RAA$, two ranges of jet \pt are selected and the corresponding jet \RAA values are plotted as a function of the average number of participants ($N_\text{part}$) in Fig.~\ref{fig:UnfoldRAA_akRadii_Npart_20_eta_20}, for jets of $80 < \pt <90$ and $130 < \pt < 150\GeVc$. The systematic uncertainty is shown in the three bounds of lines for $\RR  =  0.2$ (dotted),  0.3 (solid), and 0.4 (dashed) jets. The jet \RAA shows a clear trend of increasing suppression as the number of
participants in the PbPb collision increases.
Overall, in the kinematic range explored, the \RAA show the same level of suppression across the three distance parameters.

\begin{figure*}[htb]
  \centering
    \includegraphics[width=0.98\textwidth]{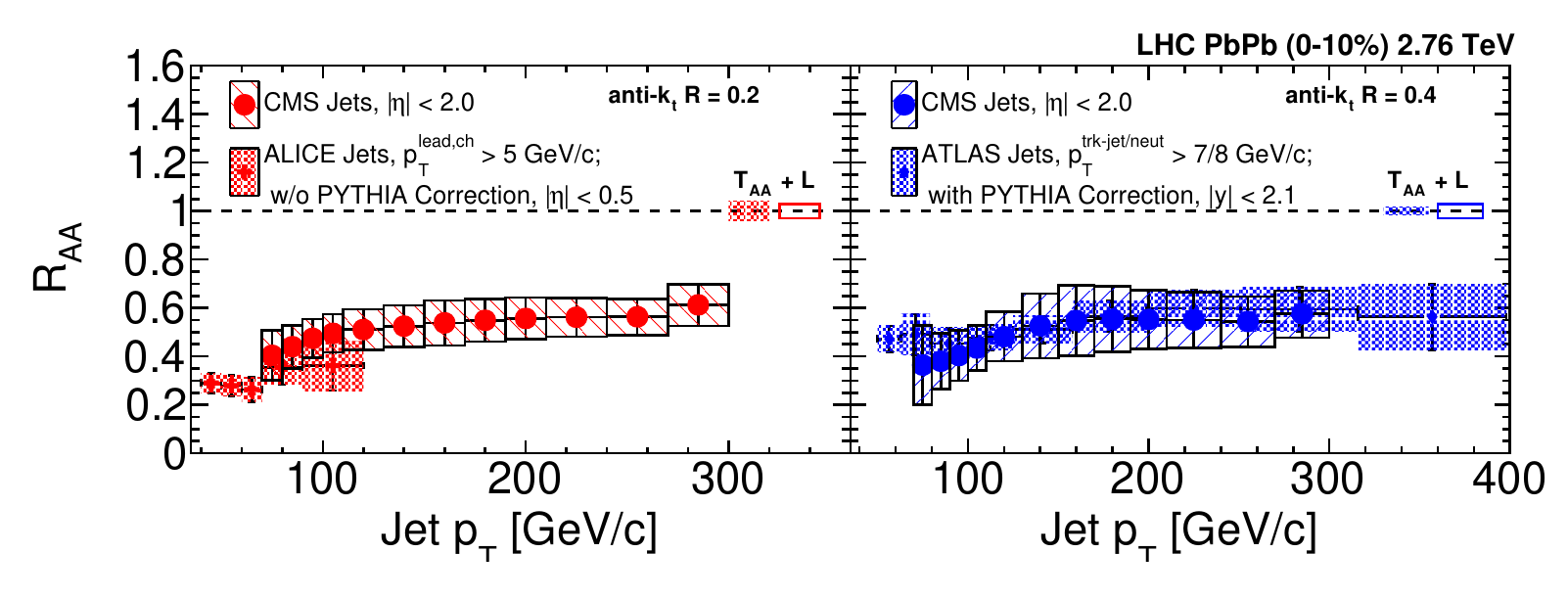}
    \caption{Left Panel: Inclusive jet \RAA as a function of the jet \pt, for \ak jets with distance parameter $\RR  =  0.2$ in the 0\%--10\% centrality bin for CMS (closed circles) and ALICE (pluses)~\cite{Adam:2015ewa}. Right Panel: Inclusive jet \RAA as a function of the jet \pt, for \ak jets with distance parameter $\RR  =  0.4$ in the 0\%--10\% centrality bin for CMS (closed circles) and ATLAS (diamonds)~\cite{Aad:2014bxa}.
The vertical bars indicate the statistical uncertainty. The systematic uncertainty is represented by the bounds of the boxes. The uncertainty boxes at unity represent the \TAA and luminosity uncertainty, open for CMS and shaded for ALICE and ATLAS. See text for a further discussion of differences in the analyses used by the three collaborations.}
    \label{fig:UnfoldRAA_CMS_ALICE_ATLAS_Npart_20_eta_20}
\end{figure*}

An experimental comparison of inclusive \ak jet \RAA  for  0-10\% centrality is shown in Fig.~\ref{fig:UnfoldRAA_CMS_ALICE_ATLAS_Npart_20_eta_20} (left panel for \ak\ jets with distance parameter $\RR  =  0.2$ for ALICE~\cite{Adam:2015ewa} and the right panel with $\RR  =  0.4$ for ATLAS~\cite{Aad:2014bxa}). Uncertainties are represented by the vertical bars for the statistical and boxes for the systematic uncertainties. The $\TAA$ and luminosity uncertainty are shown by the boxes at unity. The collection of jets for the jet \RAA calculation in these experiments differ, especially for lower jet \pt, due to the techniques employed to remove or correct the jets that did not originate in a hard scattering but that are purely due to the fluctuations in the heavy-ion underlying event. Some, but not all of the key differences are described here, for more, see ALICE~\cite{Adam:2015ewa}, ATLAS~\cite{Aad:2014bxa} and \cite{Connors:2017ptx} for a review. ALICE requires the leading track constituent of the jet to have $\pt> 5$\GeVc and constrains $\RR  =  0.2$ jets to be within $|\eta|<0.9$.  ATLAS requires its $\RR  =  0.4$ jets in $|$y$|<2.1$ to have a track jet with $\pt > 7$\GeVc or a calorimeter cluster with $\pt > 8$\GeVc within $\Delta R = 0.2$. While ALICE doesn't apply any correction on this constituent selection, ATLAS corrects for the missing jets due to this selection with correction factors estimated by {\textsc{pythia}}.  In this analysis, as described in Sec.\ref{sec:backgroundjet}, a data-driven background subtraction is introduced and all jets which are using tracks down to a $\pt$ of 0.15\GeVc and calorimeter deposits down to a $\et$ of 1\GeV are included in the jet \RAA calculation. Within the current precision of jet \RAA measurements, there is a good agreement in the overlapping \pt ranges despite the fact that the measured jet collections differ between experiments.

\section{Summary}

The cross section of \ak  particle-flow jets has been measured in pp and PbPb collisions
at $\rootsNN =2.76$\TeV for distance parameters $\RR=0.2,$ 0.3, and 0.4 in $\abs{\eta}<2$
and for jet \pt above 70\GeVc. It is found that next-to-leading order calculations with non-perturbative corrections over predict the pp cross sections, with a smaller discrepancy for larger distance parameters. The PbPb inclusive jet nuclear modification factors show a steady decrease from peripheral to central events, with a slight rise with jet \pt. No significant dependence of the jet nuclear modification factor on the distance parameter is found for the jets in the kinematic range measured in this analysis.
\clearpage
\begin{acknowledgments}
\hyphenation{Rachada-pisek}
We congratulate our colleagues in the CERN accelerator departments for the excellent performance of the LHC and thank the technical and administrative staffs at CERN and at other CMS institutes for their contributions to the success of the CMS effort. In addition, we gratefully acknowledge the computing centers and personnel of the Worldwide LHC Computing Grid for delivering so effectively the computing infrastructure essential to our analyses. Finally, we acknowledge the enduring support for the construction and operation of the LHC and the CMS detector provided by the following funding agencies: BMWFW and FWF (Austria); FNRS and FWO (Belgium); CNPq, CAPES, FAPERJ, and FAPESP (Brazil); MES (Bulgaria); CERN; CAS, MoST, and NSFC (China); COLCIENCIAS (Colombia); MSES and CSF (Croatia); RPF (Cyprus); SENESCYT (Ecuador); MoER, ERC IUT and ERDF (Estonia); Academy of Finland, MEC, and HIP (Finland); CEA and CNRS/IN2P3 (France); BMBF, DFG, and HGF (Germany); GSRT (Greece); OTKA and NIH (Hungary); DAE and DST (India); IPM (Iran); SFI (Ireland); INFN (Italy); MSIP and NRF (Republic of Korea); LAS (Lithuania); MOE and UM (Malaysia); BUAP, CINVESTAV, CONACYT, LNS, SEP, and UASLP-FAI (Mexico); MBIE (New Zealand); PAEC (Pakistan); MSHE and NSC (Poland); FCT (Portugal); JINR (Dubna); MON, RosAtom, RAS and RFBR (Russia); MESTD (Serbia); SEIDI and CPAN (Spain); Swiss Funding Agencies (Switzerland); MST (Taipei); ThEPCenter, IPST, STAR and NSTDA (Thailand); TUBITAK and TAEK (Turkey); NASU and SFFR (Ukraine); STFC (United Kingdom); DOE and NSF (USA).

Individuals have received support from the Marie-Curie program and the European Research Council and EPLANET (European Union); the Leventis Foundation; the A. P. Sloan Foundation; the Alexander von Humboldt Foundation; the Belgian Federal Science Policy Office; the Fonds pour la Formation \`a la Recherche dans l'Industrie et dans l'Agriculture (FRIA-Belgium); the Agentschap voor Innovatie door Wetenschap en Technologie (IWT-Belgium); the Ministry of Education, Youth and Sports (MEYS) of the Czech Republic; the Council of Science and Industrial Research, India; the HOMING PLUS program of the Foundation for Polish Science, cofinanced from European Union, Regional Development Fund, the Mobility Plus program of the Ministry of Science and Higher Education, the National Science Center (Poland), contracts Harmonia 2014/14/M/ST2/00428, Opus 2013/11/B/ST2/04202, 2014/13/B/ST2/02543 and 2014/15/B/ST2/03998, Sonata-bis 2012/07/E/ST2/01406; the Thalis and Aristeia programs cofinanced by EU-ESF and the Greek NSRF; the National Priorities Research Program by Qatar National Research Fund; the Programa Clar\'in-COFUND del Principado de Asturias; the Rachadapisek Sompot Fund for Postdoctoral Fellowship, Chulalongkorn University and the Chulalongkorn Academic into Its 2nd Century Project Advancement Project (Thailand); and the Welch Foundation, contract C-1845.
\end{acknowledgments}

\bibliography{auto_generated}

\cleardoublepage \appendix\section{The CMS Collaboration \label{app:collab}}\begin{sloppypar}\hyphenpenalty=5000\widowpenalty=500\clubpenalty=5000\input{HIN-13-005-authorlist.tex}\end{sloppypar}
\end{document}

%% file: HIN-13-005-authorlist.tex
\textbf{Yerevan Physics Institute,  Yerevan,  Armenia}\\*[0pt]
V.~Khachatryan, A.M.~Sirunyan, A.~Tumasyan
\vskip\cmsinstskip
\textbf{Institut f\"{u}r Hochenergiephysik,  Wien,  Austria}\\*[0pt]
W.~Adam, E.~Asilar, T.~Bergauer, J.~Brandstetter, E.~Brondolin, M.~Dragicevic, J.~Er\"{o}, M.~Flechl, M.~Friedl, R.~Fr\"{u}hwirth\cmsAuthorMark{1}, V.M.~Ghete, C.~Hartl, N.~H\"{o}rmann, J.~Hrubec, M.~Jeitler\cmsAuthorMark{1}, A.~K\"{o}nig, I.~Kr\"{a}tschmer, D.~Liko, T.~Matsushita, I.~Mikulec, D.~Rabady, N.~Rad, B.~Rahbaran, H.~Rohringer, J.~Schieck\cmsAuthorMark{1}, J.~Strauss, W.~Waltenberger, C.-E.~Wulz\cmsAuthorMark{1}
% ...
\vskip\cmsinstskip
% ...
1:~~Also at Vienna University of Technology, Vienna, Austria\\
% ...


%% file: HIN-13-005_temp.bbl
\providecommand{\href}[2]{#2}\begingroup\raggedright\begin{thebibliography}{10}%
\makeatletter
\providecommand{\hrefCMSnoop }[0]{\@secondoftwo}%
\makeatother
\providecommand{\doi}{\texttt{doi:}\begingroup \urlstyle{tt}\Url}

\bibitem{Gyulassy:1990ye}
\hrefCMSnoop {}{M.~Gyulassy and M.~Pl{\"u}mer, ``Jet quenching in dense
  matter'',} \textit{ Phys. Lett. B} \textbf{ 243} (1990) 432,
\href{http://dx.doi.org/10.1016/0370-2693(90)91409-5}{\doi{10.1016/0370-2693(90)91409-5}}.

\bibitem{Wang:1992bz}
\hrefCMSnoop {}{X.-N. Wang, ``Simulations of ultrarelativistic heavy ion
  collisions'',} in \textit{ Proceedings, 26th International Conference on
  High-energy Physics (ICHEP 92)}, p.~1812.
\newblock Dallas, Texas, USA,
August, 1992.
\newblock

\bibitem{Wiedemann:2009sh}
\hrefCMSnoop {}{U.~A. Wiedemann, ``Jet quenching in heavy ion collisions'',} in
  \textit{ Springer Materials - The Landolt-B{\"or}nstein Database}, R.~Stock,
  ed., volume 23: Relativistic Heavy Ion Physics, p.~521.
\newblock Springer-Verlag, 2010.
\newblock \href{http://www.arXiv.org/abs/0908.2306}{\texttt{arXiv:0908.2306}}.
\newblock
\href{http://dx.doi.org/10.1007/978-3-642-01539-7_17}{\doi{10.1007/978-3-642-01539-7_17}}.

\bibitem{Bjorken:1982tu}
\href {http://lss.fnal.gov/archive/1982/pub/Pub-82-059-T.pdf}{J.~D. Bjorken,
  ``Energy loss of energetic partons in {QGP}: possible extinction of high
  $\pt$ jets in hadron-hadron collisions'',} {FERMILAB-PUB-82-059-THY}, 1982.

\bibitem{Adcox:2004mh}
\hrefCMSnoop {}{{PHENIX} Collaboration, ``{Formation of dense partonic matter
  in relativistic nucleus nucleus collisions at RHIC: Experimental evaluation
  by the PHENIX collaboration}'',} \textit{ Nucl. Phys. A} \textbf{ 757} (2005)
  184,
  \href{http://dx.doi.org/10.1016/j.nuclphysa.2005.03.086}{\doi{10.1016/j.nuclphysa.2005.03.086}},
\href{http://www.arXiv.org/abs/nucl-ex/0410003}{\texttt{arXiv:nucl-ex/0410003}}.

\bibitem{Adams:2005dq}
\hrefCMSnoop {}{{STAR} Collaboration, ``{Experimental and theoretical
  challenges in the search for the quark gluon plasma: The STAR collaboration's
  critical assessment of the evidence from RHIC collisions}'',} \textit{ Nucl.
  Phys. A} \textbf{ 757} (2005) 102,
  \href{http://dx.doi.org/10.1016/j.nuclphysa.2005.03.085}{\doi{10.1016/j.nuclphysa.2005.03.085}},
\href{http://www.arXiv.org/abs/nucl-ex/0501009}{\texttt{arXiv:nucl-ex/0501009}}.

\bibitem{Back:2004je}
\hrefCMSnoop {}{{PHOBOS} Collaboration, ``{The PHOBOS perspective on
  discoveries at RHIC}'',} \textit{ Nucl. Phys. A} \textbf{ 757} (2005) 28,
  \href{http://dx.doi.org/10.1016/j.nuclphysa.2005.03.084}{\doi{10.1016/j.nuclphysa.2005.03.084}},
\href{http://www.arXiv.org/abs/nucl-ex/0410022}{\texttt{arXiv:nucl-ex/0410022}}.

\bibitem{Arsene:2004fa}
\hrefCMSnoop {}{{BRAHMS} Collaboration, ``{Quark gluon plasma and color glass
  condensate at RHIC? The perspective from the BRAHMS experiment}'',} \textit{
  Nucl. Phys. A} \textbf{ 757} (2005) 1,
  \href{http://dx.doi.org/10.1016/j.nuclphysa.2005.02.130}{\doi{10.1016/j.nuclphysa.2005.02.130}},
\href{http://www.arXiv.org/abs/nucl-ex/0410020}{\texttt{arXiv:nucl-ex/0410020}}.

\bibitem{CMS:2012aa}
\hrefCMSnoop {}{{CMS Collaboration}, ``{Study of high-$\pt$ charged particle
  suppression in PbPb compared to pp collisions at $\rootsNN=2.76$ TeV}'',}
  \textit{ Eur. Phys. J. C} \textbf{ 72} (2012) 1945,
  \href{http://dx.doi.org/10.1140/epjc/s10052-012-1945-x}{\doi{10.1140/epjc/s10052-012-1945-x}},
\href{http://www.arXiv.org/abs/1202.2554}{\texttt{arXiv:1202.2554}}.

\bibitem{Aamodt:2010jd}
\hrefCMSnoop {}{{ALICE Collaboration}, ``{Suppression of charged particle
  production at large transverse momentum in central Pb--Pb collisions at
  $\rootsNN=2.76$ TeV}'',} \textit{ Phys. Lett. B} \textbf{ 696} (2011) 30,
  \href{http://dx.doi.org/10.1016/j.physletb.2010.12.020}{\doi{10.1016/j.physletb.2010.12.020}},
\href{http://www.arXiv.org/abs/1012.1004}{\texttt{arXiv:1012.1004}}.

\bibitem{Aad:2015wga}
\hrefCMSnoop {}{{ATLAS Collaboration}, ``{Measurement of charged-particle
  spectra in Pb+Pb collisions at $\rootsNN = 2.76$ TeV with the ATLAS detector
  at the LHC}'',} \textit{ JHEP} \textbf{ 09} (2015) 050,
  \href{http://dx.doi.org/10.1007/JHEP09(2015)050}{\doi{10.1007/JHEP09(2015)050}},
\href{http://www.arXiv.org/abs/1504.04337}{\texttt{arXiv:1504.04337}}.

\bibitem{CasalderreySolana:2007zz}
\href
  {http://www.actaphys.uj.edu.pl/fulltext?series=Reg&vol=38&page=3731}{J.~Casalderrey-Solana
  and C.~A. Salgado, ``{Introductory lectures on jet quenching in heavy ion
  collisions}'',} \textit{ Acta Phys. Polon. B} \textbf{ 38} (2007) 3731,
\href{http://www.arXiv.org/abs/0712.3443}{\texttt{arXiv:0712.3443}}.

\bibitem{d'Enterria:2009am}
\hrefCMSnoop {}{D.~d'Enterria, ``Jet quenching'',} in \textit{ Springer
  Materials - The Landolt-B{\"or}nstein Database}, R.~Stock, ed., volume 23:
  Relativistic Heavy Ion Physics, p.~99.
\newblock Springer-Verlag, 2010.
\newblock \href{http://www.arXiv.org/abs/0902.2011}{\texttt{arXiv:0902.2011}}.
\newblock
\href{http://dx.doi.org/10.1007/978-3-642-01539-7_16}{\doi{10.1007/978-3-642-01539-7_16}}.

\bibitem{bass}
S.~A. Bass\hrefCMSnoop {}{ {et~al.}, ``{Systematic comparison of jet
  energy-loss schemes in a realistic hydrodynamic medium}'',} \textit{ Phys.
  Rev. C} \textbf{ 79} (2009) 024901,
  \href{http://dx.doi.org/10.1103/PhysRevC.79.024901}{\doi{10.1103/PhysRevC.79.024901}},
\href{http://www.arXiv.org/abs/0808.0908}{\texttt{arXiv:0808.0908}}.

\bibitem{Burke:2013yra}
\hrefCMSnoop {}{{JET} Collaboration, ``{Extracting the jet transport
  coefficient from jet quenching in high-energy heavy-ion collisions}'',}
  \textit{ Phys. Rev. C} \textbf{ 90} (2014) 014909,
  \href{http://dx.doi.org/10.1103/PhysRevC.90.014909}{\doi{10.1103/PhysRevC.90.014909}},
\href{http://www.arXiv.org/abs/1312.5003}{\texttt{arXiv:1312.5003}}.

\bibitem{Aad:2010bu}
\hrefCMSnoop {}{{ATLAS Collaboration}, ``{Observation of a Centrality-Dependent
  Dijet Asymmetry in Lead-Lead Collisions at $\rootsNN=2.77$ TeV with the ATLAS
  Detector at the LHC}'',} \textit{ Phys. Rev. Lett.} \textbf{ 105} (2010)
  252303,
  \href{http://dx.doi.org/10.1103/PhysRevLett.105.252303}{\doi{10.1103/PhysRevLett.105.252303}},
\href{http://www.arXiv.org/abs/1011.6182}{\texttt{arXiv:1011.6182}}.

\bibitem{Chatrchyan:2011sx}
\hrefCMSnoop {}{{CMS Collaboration}, ``{Observation and studies of jet
  quenching in PbPb collisions at nucleon-nucleon center-of-mass energy = 2.76
  TeV}'',} \textit{ Phys. Rev. C} \textbf{ 84} (2011) 024906,
  \href{http://dx.doi.org/10.1103/PhysRevC.84.024906}{\doi{10.1103/PhysRevC.84.024906}},
\href{http://www.arXiv.org/abs/1102.1957}{\texttt{arXiv:1102.1957}}.

\bibitem{Chatrchyan:2012nia}
\hrefCMSnoop {}{{CMS Collaboration}, ``{Jet momentum dependence of jet
  quenching in PbPb collisions at $\rootsNN=2.76$ TeV}'',} \textit{ Phys. Lett.
  B} \textbf{ 712} (2012) 176,
  \href{http://dx.doi.org/10.1016/j.physletb.2012.04.058}{\doi{10.1016/j.physletb.2012.04.058}},
\href{http://www.arXiv.org/abs/1202.5022}{\texttt{arXiv:1202.5022}}.

\bibitem{Chatrchyan:2013kwa}
\hrefCMSnoop {}{{CMS Collaboration}, ``{Modification of jet shapes in PbPb
  collisions at $\rootsNN = 2.76$ TeV}'',} \textit{ Phys. Lett. B} \textbf{
  730} (2014) 243,
  \href{http://dx.doi.org/10.1016/j.physletb.2014.01.042}{\doi{10.1016/j.physletb.2014.01.042}},
\href{http://www.arXiv.org/abs/1310.0878}{\texttt{arXiv:1310.0878}}.

\bibitem{Chatrchyan:2014ava}
\hrefCMSnoop {}{{CMS Collaboration}, ``{Measurement of jet fragmentation in
  PbPb and pp collisions at $\rootsNN=2.76$ TeV}'',} \textit{ Phys. Rev. C}
  \textbf{ 90} (2014) 024908,
  \href{http://dx.doi.org/10.1103/PhysRevC.90.024908}{\doi{10.1103/PhysRevC.90.024908}},
\href{http://www.arXiv.org/abs/1406.0932}{\texttt{arXiv:1406.0932}}.

\bibitem{Khachatryan:2015lha}
\hrefCMSnoop {}{{CMS Collaboration}, ``{Measurement of transverse momentum
  relative to dijet systems in PbPb and pp collisions at $ \rootsNN=2.76 $
  TeV}'',} \textit{ JHEP} \textbf{ 01} (2016) 006,
  \href{http://dx.doi.org/10.1007/JHEP01(2016)006}{\doi{10.1007/JHEP01(2016)006}},
\href{http://www.arXiv.org/abs/1509.09029}{\texttt{arXiv:1509.09029}}.

\bibitem{Khachatryan:2016erx}
\hrefCMSnoop {}{{CMS Collaboration}, ``{Correlations between jets and charged
  particles in PbPb and pp collisions at $\rootsNN=2.76 $ TeV}'',} \textit{
  JHEP} \textbf{ 02} (2016) 156,
  \href{http://dx.doi.org/10.1007/JHEP02(2016)156}{\doi{10.1007/JHEP02(2016)156}},
\href{http://www.arXiv.org/abs/1601.00079}{\texttt{arXiv:1601.00079}}.

\bibitem{majumder}
\hrefCMSnoop {}{A.~Majumder, ``{A comparative study of jet-quenching
  schemes}'',} \textit{ J. Phys. G} \textbf{ 34} (2007) S377,
  \href{http://dx.doi.org/10.1088/0954-3899/34/8/S25}{\doi{10.1088/0954-3899/34/8/S25}},
\href{http://www.arXiv.org/abs/nucl-th/0702066}{\texttt{arXiv:nucl-th/0702066}}.

\bibitem{Casalderrey-Solana:2015bww}
\hrefCMSnoop {}{J.~Casalderrey-Solana, D.~Pablos, and K.~Tywoniuk, ``Two-gluon
  emission and interference in a thin {QCD} medium: insights into jet
  formation'',} \textit{ JHEP} \textbf{ 11} (2016) 174,
  \href{http://dx.doi.org/10.1007/JHEP11(2016)174}{\doi{10.1007/JHEP11(2016)174}},
\href{http://www.arXiv.org/abs/1512.07561}{\texttt{arXiv:1512.07561}}.

\bibitem{Chien:2015hda}
\hrefCMSnoop {}{Y.-T. Chien and I.~Vitev, ``{Towards the understanding of jet
  shapes and cross sections in heavy ion collisions using soft-collinear
  effective theory}'',} \textit{ JHEP} \textbf{ 05} (2016) 023,
  \href{http://dx.doi.org/10.1007/JHEP05(2016)023}{\doi{10.1007/JHEP05(2016)023}},
\href{http://www.arXiv.org/abs/1509.07257}{\texttt{arXiv:1509.07257}}.

\bibitem{Aad:2014bxa}
\hrefCMSnoop {}{{ATLAS Collaboration}, ``{Measurements of the Nuclear
  Modification Factor for Jets in Pb+Pb Collisions at $\rootsNN=2.76$ TeV with
  the ATLAS Detector}'',} \textit{ Phys. Rev. Lett.} \textbf{ 114} (2015)
  072302,
  \href{http://dx.doi.org/10.1103/PhysRevLett.114.072302}{\doi{10.1103/PhysRevLett.114.072302}},
\href{http://www.arXiv.org/abs/1411.2357}{\texttt{arXiv:1411.2357}}.

\bibitem{Adam:2015ewa}
\hrefCMSnoop {}{{ALICE Collaboration}, ``{Measurement of jet suppression in
  central Pb-Pb collisions at $\rootsNN=2.76$ TeV}'',} \textit{ Phys. Lett. B}
  \textbf{ 746} (2015) 1,
  \href{http://dx.doi.org/10.1016/j.physletb.2015.04.039}{\doi{10.1016/j.physletb.2015.04.039}},
\href{http://www.arXiv.org/abs/1502.01689}{\texttt{arXiv:1502.01689}}.

\bibitem{Khachatryan:2016xdg}
\hrefCMSnoop {}{{CMS Collaboration}, ``{Measurement of inclusive jet production
  and nuclear modifications in pPb collisions at $\rootsNN$ = 5.02 TeV}'',}
  \textit{ Eur. Phys. J. C} \textbf{ 76} (2016) 372,
  \href{http://dx.doi.org/10.1140/epjc/s10052-016-4205-7}{\doi{10.1140/epjc/s10052-016-4205-7}},
\href{http://www.arXiv.org/abs/1601.02001}{\texttt{arXiv:1601.02001}}.

\bibitem{Adam:2015hoa}
\hrefCMSnoop {}{{ALICE Collaboration}, ``{Measurement of charged jet production
  cross sections and nuclear modification in p-Pb collisions at $\rootsNN =
  5.02$ TeV}'',} \textit{ Phys. Lett. B} \textbf{ 749} (2015) 68,
  \href{http://dx.doi.org/10.1016/j.physletb.2015.07.054}{\doi{10.1016/j.physletb.2015.07.054}},
\href{http://www.arXiv.org/abs/1503.00681}{\texttt{arXiv:1503.00681}}.

\bibitem{ATLAS:2014cpa}
\hrefCMSnoop {}{{ATLAS Collaboration}, ``{Centrality and rapidity dependence of
  inclusive jet production in $\rootsNN = 5.02$ TeV proton-lead collisions with
  the ATLAS detector}'',} \textit{ Phys. Lett. B} \textbf{ 748} (2015) 392,
  \href{http://dx.doi.org/10.1016/j.physletb.2015.07.023}{\doi{10.1016/j.physletb.2015.07.023}},
\href{http://www.arXiv.org/abs/1412.4092}{\texttt{arXiv:1412.4092}}.

\bibitem{Cacciari:2011ma}
\hrefCMSnoop {}{M.~Cacciari, G.~P. Salam, and G.~Soyez, ``{FastJet user
  manual}'',} \textit{ Eur. Phys. J. C} \textbf{ 72} (2012) 1896,
  \href{http://dx.doi.org/10.1140/epjc/s10052-012-1896-2}{\doi{10.1140/epjc/s10052-012-1896-2}},
\href{http://www.arXiv.org/abs/1111.6097}{\texttt{arXiv:1111.6097}}.

\bibitem{Miller:2007ri}
\hrefCMSnoop {}{M.~L. Miller, K.~Reygers, S.~J. Sanders, and P.~Steinberg,
  ``{Glauber modeling in high energy nuclear collisions}'',} \textit{ Ann. Rev.
  Nucl. Part. Sci.} \textbf{ 57} (2007) 205,
  \href{http://dx.doi.org/10.1146/annurev.nucl.57.090506.123020}{\doi{10.1146/annurev.nucl.57.090506.123020}},
\href{http://www.arXiv.org/abs/nucl-ex/0701025}{\texttt{arXiv:nucl-ex/0701025}}.

\bibitem{Chatrchyan:2008aa}
\hrefCMSnoop {}{{CMS Collaboration}, ``{The CMS experiment at the CERN LHC}'',}
  \textit{ JINST} \textbf{ 3} (2008) S08004,
\href{http://dx.doi.org/10.1088/1748-0221/3/08/S08004}{\doi{10.1088/1748-0221/3/08/S08004}}.

\bibitem{CMS-PAS-JME-07-003}
\href {http://cdsweb.cern.ch/record/1198227}{{CMS Collaboration}, ``Performance
  of jet algorithms in {CMS}'',} CMS Physics Analysis Summary
  CMS-PAS-JME-07-003, 2007.

\bibitem{Kodolova:2007hd}
\hrefCMSnoop {}{O.~Kodolova, I.~Vardanian, A.~Nikitenko, and A.~Oulianov,
  ``{The performance of the jet identification and reconstruction in heavy ions
  collisions with CMS detector}'',} \textit{ Eur. Phys. J. C} \textbf{ 50}
  (2007) 117,
\href{http://dx.doi.org/10.1140/epjc/s10052-007-0223-9}{\doi{10.1140/epjc/s10052-007-0223-9}}.

\bibitem{bib_pythia}
\hrefCMSnoop {}{T.~Sj{\"o}strand, S.~Mrenna, and P.~Skands, ``{PYTHIA} 6.4
  physics and manual'',} \textit{ JHEP} \textbf{ 05} (2006) 026,
  \href{http://dx.doi.org/10.1088/1126-6708/2006/05/026}{\doi{10.1088/1126-6708/2006/05/026}},
\href{http://www.arXiv.org/abs/hep-ph/0603175}{\texttt{arXiv:hep-ph/0603175}}.

\bibitem{Lokhtin:2005px}
\hrefCMSnoop {}{I.~P. Lokhtin and A.~M. Snigirev, ``{A model of jet quenching
  in ultrarelativistic heavy ion collisions and high-p$_{T}$ hadron spectra at
  RHIC}'',} \textit{ Eur. Phys. J. C} \textbf{ 45} (2006) 211,
  \href{http://dx.doi.org/10.1140/epjc/s2005-02426-3}{\doi{10.1140/epjc/s2005-02426-3}},
\href{http://www.arXiv.org/abs/hep-ph/0506189}{\texttt{arXiv:hep-ph/0506189}}.

\bibitem{Chatrchyan:2012gt}
\hrefCMSnoop {}{{CMS Collaboration}, ``{Studies of jet quenching using
  isolated-photon+jet correlations in PbPb and pp collisions at $\rootsNN=2.76$
  TeV}'',} \textit{ Phys. Lett. B} \textbf{ 718} (2013) 773,
  \href{http://dx.doi.org/10.1016/j.physletb.2012.11.003}{\doi{10.1016/j.physletb.2012.11.003}},
\href{http://www.arXiv.org/abs/1205.0206}{\texttt{arXiv:1205.0206}}.

\bibitem{CMS:2009nxa}
\href {https://cds.cern.ch/record/1194487}{{CMS Collaboration}, ``Particle-flow
  event reconstruction in cms and performance for jets, taus, and met'',} CMS
  Physics Analysis Summary CMS-PAS-PFT-09-001, 2009.

\bibitem{CMS:2010byl}
\href {https://cds.cern.ch/record/1247373}{{CMS Collaboration},
  ``{Commissioning of the Particle-flow Event Reconstruction with the first LHC
  collisions recorded in the CMS detector}'',} CMS Physics Analysis Summary
  CMS-PAS-PFT-10-001, 2010.

\bibitem{Chatrchyan:2011ds}
\hrefCMSnoop {}{{CMS Collaboration}, ``Determination of jet energy calibration
  and transverse momentum resolution in {CMS}'',} \textit{ JINST} \textbf{ 6}
  (2011) P11002,
  \href{http://dx.doi.org/10.1088/1748-0221/6/11/P11002}{\doi{10.1088/1748-0221/6/11/P11002}},
\href{http://www.arXiv.org/abs/1107.4277}{\texttt{arXiv:1107.4277}}.

\bibitem{Elden:1982wp}
\hrefCMSnoop {}{L.~Eld\'en, ``A weighted pseudoinverse, generalized singular
  values, and constrained least squares problems'',} \textit{ BIT Numer. Math.}
  \textbf{ 22} (1982) 487,
  \href{http://dx.doi.org/10.1007/BF01934412}{\doi{10.1007/BF01934412}}.

\bibitem{D'Agostini:1994zf}
\hrefCMSnoop {}{G.~D'Agostini, ``A multidimensional unfolding method based on
  {Bayes'} theorem'',} \textit{ Nucl. Instrum. Meth. A} \textbf{ 362} (1995)
  487,
\href{http://dx.doi.org/10.1016/0168-9002(95)00274-X}{\doi{10.1016/0168-9002(95)00274-X}}.

\bibitem{Hansen:1998rd}
P.~C. Hansen, ``Rank-Deficient and Discrete Ill-Posed Problems: Numerical
  Aspects of Linear Inversion''.
\newblock SIAM, Philadelphia, PA, 1998.
\newblock
  \href{http://dx.doi.org/10.1137/1.9780898719697}{\doi{10.1137/1.9780898719697}},
  ISBN~978-0898714036.

\bibitem{Adye:2011gm}
\hrefCMSnoop {}{T.~Adye, ``{Unfolding algorithms and tests using RooUnfold}'',}
  in \textit{ {Proceedings, PHYSTAT 2011 Workshop on Statistical Issues Related
  to Discovery Claims in Search Experiments and Unfolding, CERN,Geneva,
  Switzerland 17-20 January 2011}}, pp.~313--318, CERN.
\newblock CERN, Geneva, 2011.
\newblock \href{http://www.arXiv.org/abs/1105.1160}{\texttt{arXiv:1105.1160}}.
\newblock
\href{http://dx.doi.org/10.5170/CERN-2011-006.313}{\doi{10.5170/CERN-2011-006.313}}.

\bibitem{CMS-PAS-LUM-13-002}
\href {http://cdsweb.cern.ch/record/1643269}{{CMS Collaboration}, ``Luminosity
  calibration for the 2013 proton-lead and proton-proton data taking'',} CMS
  Physics Analysis Summary CMS-PAS-LUM-13-002, 2013.

\bibitem{Wobisch:2011ij}
\hrefCMSnoop {}{{fastNLO} Collaboration, ``{Theory-Data Comparisons for Jet
  Measurements in Hadron-Induced Processes}'',} (2011).
\href{http://www.arXiv.org/abs/1109.1310}{\texttt{arXiv:1109.1310}}.

\bibitem{Ball:2012cx}
\hrefCMSnoop {}{{NNPDF} Collaboration, ``{Parton distributions with LHC
  data}'',} \textit{ Nucl. Phys. B} \textbf{ 867} (2013) 244,
  \href{http://dx.doi.org/10.1016/j.nuclphysb.2012.10.003}{\doi{10.1016/j.nuclphysb.2012.10.003}},
\href{http://www.arXiv.org/abs/1207.1303}{\texttt{arXiv:1207.1303}}.

\bibitem{Gao:2013xoa}
J.~Gao\hrefCMSnoop {}{ {et~al.}, ``{CT10 next-to-next-to-leading order global
  analysis of QCD}'',} \textit{ Phys. Rev. D} \textbf{ 89} (2014) 033009,
  \href{http://dx.doi.org/10.1103/PhysRevD.89.033009}{\doi{10.1103/PhysRevD.89.033009}},
\href{http://www.arXiv.org/abs/1302.6246}{\texttt{arXiv:1302.6246}}.

\bibitem{Khachatryan:2015luy}
\hrefCMSnoop {}{{CMS Collaboration}, ``{Measurement of the inclusive jet cross
  section in pp collisions at $\sqrt{s} = 2.76\,\text {TeV}$}'',} \textit{ Eur.
  Phys. J. C} \textbf{ 76} (2016) 265,
  \href{http://dx.doi.org/10.1140/epjc/s10052-016-4083-z}{\doi{10.1140/epjc/s10052-016-4083-z}},
\href{http://www.arXiv.org/abs/1512.06212}{\texttt{arXiv:1512.06212}}.

\bibitem{Connors:2017ptx}
\hrefCMSnoop {}{M.~Connors, C.~Nattrass, R.~Reed, and S.~Salur, ``{Review of
  Jet Measurements in Heavy Ion Collisions}'',}
\href{http://www.arXiv.org/abs/1705.01974}{\texttt{arXiv:1705.01974}}.

\end{thebibliography}\endgroup
